\documentclass[aps,prd,nobibnotes,twocolumn,groupedaddress,preprintnumbers,showpacs,showkeys,tightenlines,superscriptaddress,nofootinbib]{revtex4-2}
\usepackage{lipsum}
\usepackage{hyperref}
\usepackage{epsfig}
\usepackage{latexsym}
\usepackage{amssymb}
\usepackage{booktabs}
\usepackage{graphicx}\usepackage{physics}
\usepackage{simpler-wick}

\usepackage{xcolor}
\usepackage{amsmath}
\usepackage{bm}

\let\oldeqref\eqref
\renewcommand{\eqref}[1]{Eq.~\oldeqref{#1}}

\newcommand{\figref}[1]{Fig.~\ref{#1}}
\newcommand{\tabref}[1]{Table~\ref{#1}}
\newcommand{\tabsref}[2]{Tables.~\ref{#1}-~\ref{#2}}
\newcommand{\secref}[1]{Sec.~\ref{#1}}
\newcommand{\cov}{\text{\textbf{cov}}}

\begin{document}
\preprint{MITP-25-026}
\title{The isoscalar octet axial form factor of the nucleon from lattice QCD}

\author{Alessandro Barone}
\email{abarone@uni-mainz.de}
\affiliation{PRISMA$^+$ Cluster of Excellence \& Institut f\"ur Kernphysik,
 Johannes Gutenberg-Universit\"at  Mainz,  D-55099 Mainz, Germany}
 
\author{Dalibor~Djukanovic} 
\affiliation{Helmholtz-Institut Mainz, Johannes Gutenberg-Universit\"at Mainz,
D-55099 Mainz, Germany}
 \affiliation{GSI Helmholtzzentrum für Schwerionenforschung, 64291 Darmstadt, Germany}

\author{Georg~von~Hippel}
\affiliation{PRISMA$^+$ Cluster of Excellence \& Institut f\"ur Kernphysik,
 Johannes Gutenberg-Universit\"at  Mainz,  D-55099 Mainz, Germany}

\author{Jonna~Koponen}
\affiliation{PRISMA$^+$ Cluster of Excellence  \& Institut f\"ur Kernphysik,
Johannes Gutenberg-Universit\"at Mainz, D-55099 Mainz, Germany}

\author{Harvey~B.~Meyer} 
\affiliation{PRISMA$^+$ Cluster of Excellence  \& Institut f\"ur Kernphysik,
Johannes Gutenberg-Universit\"at Mainz, D-55099 Mainz, Germany}
\affiliation{Helmholtz-Institut Mainz, Johannes Gutenberg-Universit\"at Mainz,
D-55099 Mainz, Germany}

\author{Konstantin~Ottnad}
\affiliation{PRISMA$^+$ Cluster of Excellence \& Institut f\"ur Kernphysik,
 Johannes Gutenberg-Universit\"at  Mainz,  D-55099 Mainz, Germany}

\author{Hartmut~Wittig} 
 \affiliation{PRISMA$^+$ Cluster of Excellence  \& Institut f\"ur Kernphysik,
Johannes Gutenberg-Universit\"at Mainz, D-55099 Mainz, Germany}
\affiliation{Helmholtz-Institut Mainz, Johannes Gutenberg-Universit\"at Mainz,
D-55099 Mainz, Germany}

\begin{abstract}
The isoscalar axial-vector form factor of the nucleon plays a key role in
	understanding the electroweak interaction of nucleons. For the
	interpretation of the spin structure of the nucleon the non-singlet
	isoscalar axial charge is indispensable. Moreover, $G_A^{u+d-2s}(Q^2)$ together
	with the isovector and singlet isoscalar form factors are needed for
	the flavor decomposition of the axial-vector form factor. Detailed
	knowledge of the flavor decomposition facilitates extractions of
	Standard Model (SM) parameters from low energy experiments such as the
	electroweak charge of the proton in the P2 experiment. Here we present
	a lattice determination of $G_A^{u+d-2s}(Q^2)$ on $N_f=2+1$ $\mathcal{O}(a)$
	improved Wilson fermions, with a full error budget concerning extrapolations and
	interpolations to the continuum and infinite volume and physical quark
	masses.

\end{abstract}

\maketitle

\section{Introduction\label{sec:intro}}
The axial form factor of the nucleon is crucial for the understanding of the electroweak
interaction of nucleons, mediated by $W$ or $Z$ boson exchange. In  the isospin
symmetric case the $W$ boson exchange is sensitive to the $u-d$ flavor combination
of the axial current, while the $Z$ boson exchange also receives contributions
from the strange quark. The latter enters prominently in asymmetry measurements of parity
violating electron-proton scattering, which allows for the determination of the weak charge
of the proton \cite{Maas:2017snj}. The isovector axial-vector form factor of
the nucleon is instrumental for the analysis of upcoming neutrino experiments
\cite{DUNE:2015lol,Hyper-Kamiokande:2018ofw}, with a recent increased effort
from lattice QCD
\cite{RQCD:2019jai,Hasan:2019noy,Alexandrou:2020okk,Gupta:2017dwj,Djukanovic:2022wru,Jang:2019vkm,Tomalak:2023pdi,Jang:2023zts}
to calculate the form factor with the precision necessary to facilitate the
interpretation of experimental results. 

A determination of the octet axial-vector charge, i.e. the flavor combination
$u+d-2s$, is crucial in order to constrain the flavor structure of the proton
spin decomposition at the quark level, e.g.  based on data from polarized deep
inelastic scattering \cite{Bass:2009ed}.  Assuming SU(3) flavor symmetry, the
axial-vector charge for the octet combination can be obtained from semileptonic
beta decay of hyperons \cite{Close:1993mv}. However, SU(3) may be badly broken
\cite{Savage:1996zd},
and in baryon chiral perturbation theory (BChPT) at next-to-leading-order \cite{Ledwig:2014rfa}
the charge receives a contribution from an unknown SU(3)-breaking low-energy
constant, which may not be extracted from semileptonic hyperon decays.
Thus, a non-perturbative determination of the octet axial-vector
charge is necessary to reduce the model dependence when interpreting
polarized deep inelastic scattering data.
Moreover, the octet combination is another step towards the full flavor
decomposition of the axial-vector form factors, needed e.g. for the precise
extraction of the weak charge of the proton in the P2 experiment
\cite{Becker:2018ggl}.
The full flavor decomposition relies on the computation of both octet and singlet contributions. However,
these renormalize differently: we therefore focus only on the former, leaving the computation of the appropriate
renormalization factor and form factor for the latter for future studies.
For a recent review on the status of the spin
decomposition as obtained from lattice QCD see Ref.~\cite{Liu:2021lke}.

Here we present the determination of the axial-vector
octet combination form factor from lattice QCD, where (in Minkowskian notation) we use the
following parameterization for the flavor diagonal currents,
\begin{align}
	\nonumber
	\langle N_{\rm p^+} | A_\mu^a
	|N_{\rm p^+}\rangle& = \bar u (p')\Bigl [ G_A^a(Q^2) \gamma_\mu +
	\frac{q_\mu}{2m} G_P^a(Q^2)\Bigr] \gamma_5\,u(p), \\
\label{eq:defcurrent}
	A_\mu^a&= \bar \psi \ T^a   \gamma_\mu \gamma_5 \
	 \psi, \quad \ a \in \{0,3,8\}, \phantom{{\bigg|}}\\ \nonumber
	T^0 &= 
	\left(\begin{array}{ccc}
		1 & 0 & 0\\
		0 & 1 & 0\\
		0 & 0 & 1
	\end{array}
		\right),~~
	T^3 = 
	\left(\begin{array}{ccc}
		1 & 0 & 0\\
		0 & -1 & 0\\
		0 & 0 & 0
	\end{array}
		\right),\nonumber\\
	T^8&= 
	\left(\begin{array}{ccc}
		1 & 0 & 0\\
		0 & 1 & 0\\
		0 & 0 & -2
	\end{array}
		\right),~~
	\psi= 
	\left(\begin{array}{c}
		u\\
		d\\
		s
	\end{array}
		\right),\nonumber
\end{align} 
where $p$ $(p')$ is the momentum of the incoming (outgoing) proton, $q=p'-p$
the momentum transfer, $p^2=p'{}^2=m^2$ and $q^2= -Q^2$. Since we
will only be concerned with the octet flavor combination we will drop the
superscript in the following. The calculation follows closely the procedures
presented in Ref.~\cite{Djukanovic:2022wru} for the connected part of the
calculation.  For the disconnected contributions we use the same techniques
applied in Ref.~\cite{Agadjanov:2023efe}.

\section{Setup\label{sec:setup}}
The calculation is based on $N_f=2+1$ lattice simulations with $\mathcal{O}(a)$
improved Wilson fermions \cite{Sheikholeslami:1985ij} and tree-level improved L{\"u}scher-Weisz
gauge action \cite{Luscher:1984xn},
performed within the coordinated lattice simulation (CLS) effort
\cite{Bruno:2014jqa}.
We perform the necessary reweighting for the twisted-mass
and rational approximation using \cite{Mohler:2020txx,Kuberski:2023zky}.
The set of ensembles used in this work is identical to \cite{Djukanovic:2022wru},
and a summary is given in~\tabref{tab:ensembles}.

\begin{figure}[t]
	\includegraphics{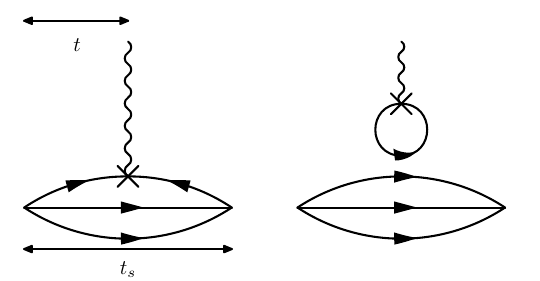}
	\caption{Wick contractions of the axial current $A^a_\mu$ of
	Eq.~(\ref{eq:defcurrent}), denoted by a
	cross, with the
	nucleon interpolating operator of Eq.~(\ref{eq:interp}). Left: the quark-line
	connected contribution, where $t$ denotes the distance between source
	time slice and operator insertion time slice, and $t_s$
	denotes the distance between source and sink time slice, respecetively. Right:
	the quark-line disconnected diagram.}
\label{fig:diagram}
\end{figure}

\begin{figure}
\centering
\includegraphics[scale=0.25]{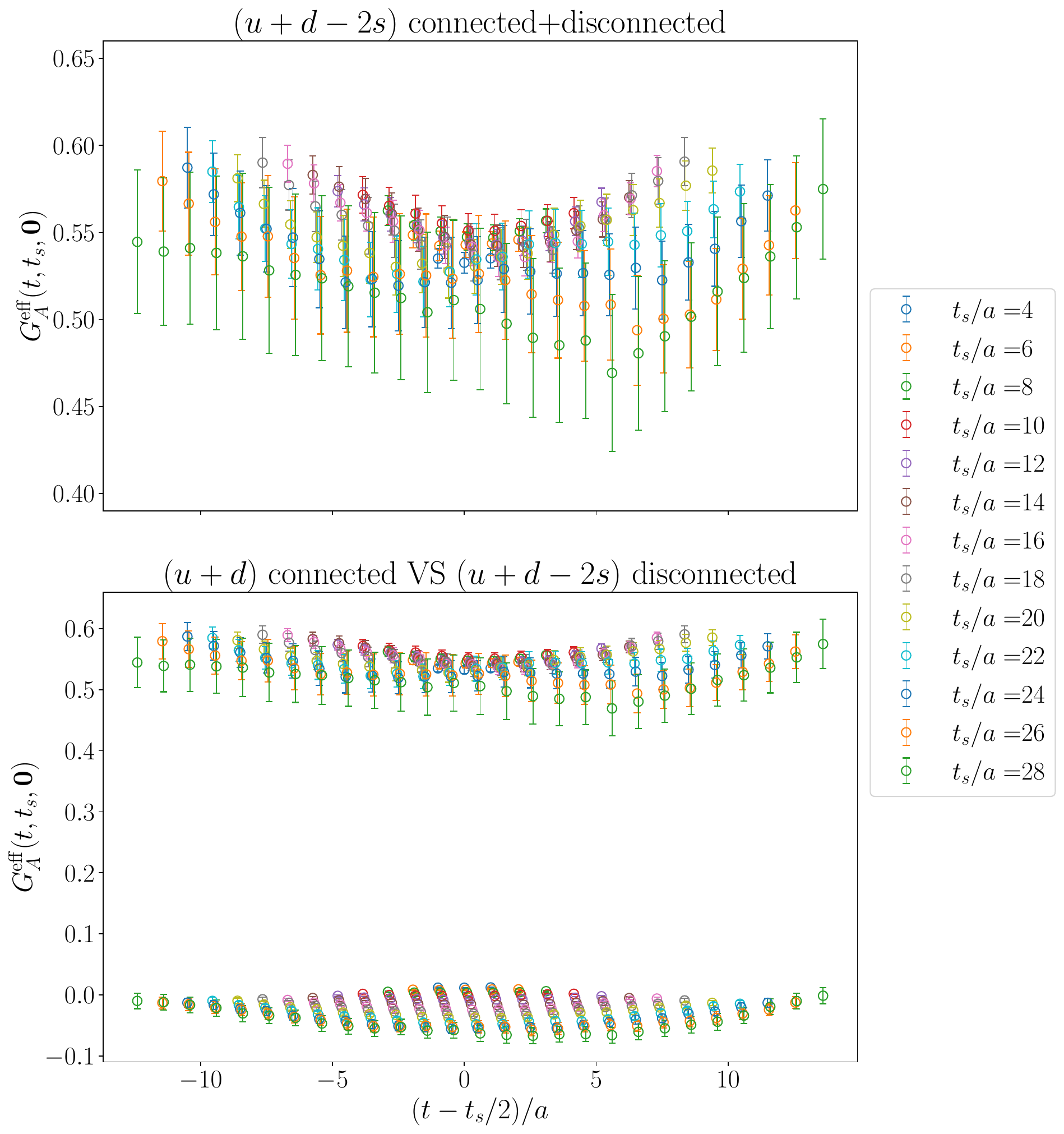}
\caption{Effective form factor $G_A^{\rm eff}$ extracted from the ratio~\eqref{eq:ratio} at vanishing $Q^2$ on the ensemble E300 for all the available source-sink separations $t_s$.
The upper plot shows the full contribution to the octet combination, whereas the lower plot shows the breakdown into connected pieces (upper points) and
disconnected pieces (lower points).
The plots already include the relevant renormalization factors.}
\label{fig:GAeff_connVSdisc}
\end{figure}

\begin{figure}
\centering
\includegraphics[scale=0.3]{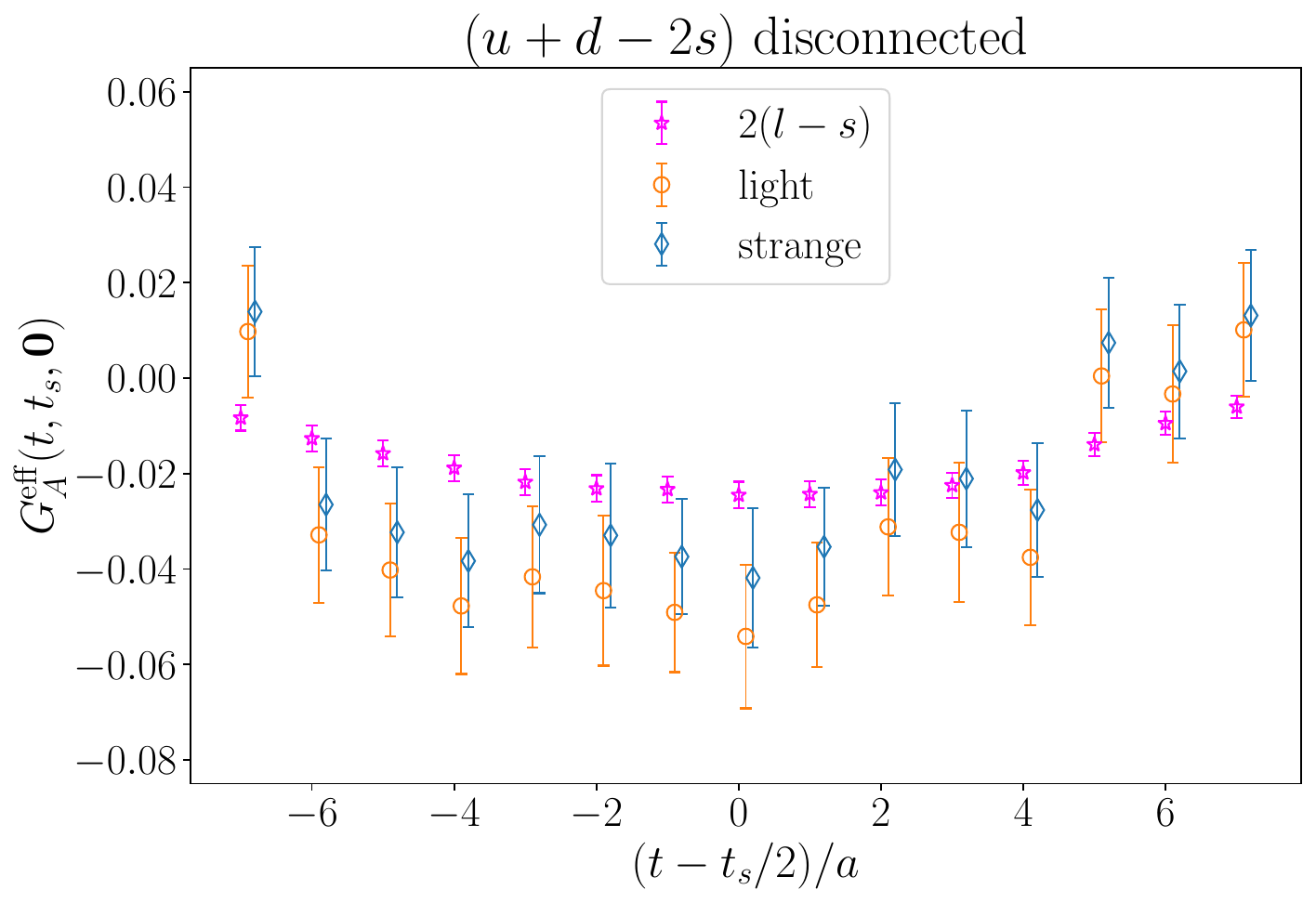}
\caption{Effective form factor $G_A^{\rm eff}$ extracted from the ratio~\eqref{eq:ratio} at vanishing $Q^2$ on the ensemble E300 at $t_s=\nobreak0.79\,\text{fm}$
for the light ($l$) and strange ($s$) disconnected contributions separately.
The plot already includes the relevant renormalization factors for the octet contributions.}
\label{fig:GAeff_disc}
\end{figure}

For the extraction of the axial-vector form factor $G_A(Q^2)$ of \eqref{eq:defcurrent}, the
pertinent Wick contractions lead to quark-line connected and quark-line disconnected
contributions (cf.~\figref{fig:diagram}).
The calculation of the connected part follows the strategy from~\cite{Djukanovic:2022wru}:
the number of sources, the available source-sink separations and the computational setup are identical.
We first calculate the two- and three-point
functions, which read
\begin{align}
	C_2^\Gamma(t,\bf{p})\hspace{-0.1em}&=\hspace{-0.1em}\Gamma_{\alpha \beta} \hspace{-0.2em}\sum\limits_{\bf{x}} e^{-i {\bf p \cdot x }}
	\langle \Psi_\beta(\mathbf{x},t) \overline{\Psi}_\alpha (0)\rangle,\\ \hspace{-0.3cm}
	C^{\Gamma'}_3(t,t_{s},{\bf q}) \hspace{-0.1em} &= \hspace{-0.1em}\Gamma'_{\alpha\beta} \hspace{-0.2em} \sum\limits_{\bf
	x, y} e^{i {\bf q \cdot y}} \langle \Psi_\beta(\mathbf{x},t_s) A^8_\mu(\mathbf{y},t)\,
	\overline{\Psi}_\alpha (0)\rangle,  
\end{align}
where the interpolating operator of the nucleon reads
\begin{align}
	\Psi_\alpha (x) = \varepsilon_{abc} \Bigl( \tilde{u} _a^T (x) C \gamma_5
	\tilde{d}_b(x) \Bigr) \tilde{u}_{c,\alpha} (x), \label{eq:interp}
\end{align}
built from Gaussian smeared quark fields \cite{Gusken:1989ad}
\begin{align}
 \tilde{q} = (1+\kappa_G\triangle )^{N_G} q
\end{align}
using spatially APE-smeared gauge links in the covariant Laplacian
\cite{APE:1987ehd}.
We then form the ratio
\begin{align}
	R(t,t_s,{\bf q} )& =\frac{C^{\Gamma}_3(t,t_s,{\bf
	q})}{C^{\Gamma}_2(t_s,{\bf 0})} \nonumber \\
	&\times \sqrt{
		\frac{C^{\Gamma}_2(t_s -t ,{\bf q}) C^{\Gamma}_2 (t,{\bf{0}})
		C^{\Gamma}_2(t_s,{\bf 0})}{C^{\Gamma}_2(t_{s}-t,{\bf 0})
		C^{\Gamma}_2(t,{\bf q}) C^{\Gamma}_2(t_s,{\bf q})} 
	} \label{eq:ratio}
\end{align}
which in the limit of large source-sink separations allows for the extraction of
the axial form factor $G_A(Q^2)$.
The calculation of the disconnected part is described in~\cite{Agadjanov:2023efe}: we
first calculate
the one-point functions 
\begin{align}
	\mathcal L (z_0,{\bf q})&= -\sum\limits_{\bf z} e^{i {\bf q\cdot z }
	} \ \mathrm{Tr} \Bigl[ S_q^{-1}(z,z) \gamma_\mu \gamma_5\Bigr],
\end{align}
where $S_q^{-1}$ is the propagator for the quark $q$,
and then build the disconnected three-point functions
\begin{align}
	C_3^{\mathrm{disc},\Gamma'} (t,t_s,{\bf q}) &=
\langle \mathcal{L} (t,{\bf q })\ \mathcal{C}^{\Gamma'}_2(t_{s},{\bf
	0})\rangle,
\label{eq:3pt_disc}
\end{align}
where $\mathcal{C}_2$ denotes the Wick contraction of the interpolating operators
of Eq.~(\ref{eq:interp}),
\begin{align}
	\mathcal{C}^{\Gamma'}_2 (t_s,{\bf p'}) &= \sum_{\bf x} e^{-i {\bf p'\cdot  x}} \, 
	\text{Tr} \Bigl[ \Gamma' \cdot \wick{\c1 \Psi(\mathbf{x},t_s) \, {\overline{\c1
	\Psi}(0)} }
	\Bigr],
\end{align}
and the traces are understood in Dirac space.
For the connected part we always choose the polarization matrix $\Gamma =
\frac{1}{2} (1+\gamma_0) (1+i \gamma_5 \gamma_3)$, whereas for  the disconnected
we average over all three different polarizations, i.e. 
\begin{align}
	\Gamma'_i= \frac{1}{2}(1+\gamma_0) (1+ i \gamma_5 \gamma_i),\qquad
	i=1,2,3.
\end{align}
The disconnected diagrams are notoriously difficult to calculate precisely, and
consequently we increased the statistics for the
two-point functions entering~\eqref{eq:3pt_disc} with respect to the number
of sources of the two-point correlators entering
the ratio~\eqref{eq:ratio} for the connected pieces.
When combining the connected
and disconnected part, we match the analyzed source-sink separations $t_s$ of the disconnected
to the setup of the connected diagrams. The effective form factors are
then calculated separately for the connected and disconnected parts. Nevertheless, we
perform a jackknife analysis to preserve correlations among the configurations
of each ensemble.
\begin{table}[t]
	\begin{tabular}{cccccc}
\toprule
		ID & $\beta$ & $T/a$ & $L/a$ & $M_\pi$ [MeV] & $M_N$ MeV
		\\\hline
		H102 & 3.40 & 96 & 32 & 354 & 1103 \\
		H105 & 3.40 & 96 & 32 & 280 & 1045 \\
		C101 & 3.40 & 96 & 48 & 225 & 980 \\
		N101 & 3.40 & 128 & 48 & 281 & 1030\\\hline
		S400 & 3.46 & 128 & 32 & 350 & 1130 \\
		N451 & 3.46 & 128 & 48 & 286 & 1011\\
		D450 & 3.46 & 128 & 64 & 216 & 978\\ \hline
		N203 & 3.55 & 128 & 48 & 346 & 1112\\
		N200 & 3.55 & 128 & 48 & 281 & 1063\\
		D200 & 3.55 & 128 & 64 & 203 & 966 \\
		E250 & 3.55 & 192 & 96 & 129 & 928 \\\hline
		N302 & 3.70 & 128 & 48 & 348 & 1146 \\
		J303 & 3.70 & 192 & 64 & 260 & 1048 \\
		E300 & 3.70 & 192 & 96 & 174 & 962 \\
		\bottomrule
	\end{tabular}
	\caption{Summary of ensembles used, where the values for
	$\beta=3.40,3.46,3.55,3.70$ correspond to a lattice spacing of roughly
	$a\sim 0.086,0.076,0.064,0.050$ fm, respectively~\cite{PhysRevD.95.074504}. }
	\label{tab:ensembles}
\end{table}
For the connected piece we use the strategy of \cite{Djukanovic:2022wru}, where
the spatial momentum combinations and the component of the axial current are chosen
such that the ratio of Eq.~(\ref{eq:ratio}) is only sensitive to $G_A(Q^2)$ and
automatically $\mathcal{O}(a)$ improved. For the disconnected pieces we
choose to solve the system of equations as outlined in \cite{Capitani:2017qpc},
restricting the current to its spatial components only. We show an example of the raw data
for vanishing momentum on the ensemble E300 in~\figref{fig:GAeff_connVSdisc}. This illustrates the quality of the signal as well as the magnitude
of both connected and disconnected contributions. 
We emphasize that the signal relies on the difference $2(l-s)$, where $l$ refers to the light quark,
and it benefits from the correlations between the two flavors, as shown in~\figref{fig:GAeff_disc}.

The renormalization
pattern of the improved operators is given in Ref.~\cite{Bhattacharya:2005rb},
where the improved operators read
\begin{align}
	A^{a,I}_\mu(x)=\Bigl (A^a_\mu(x) +a \ c_A \tilde{\partial}_\mu
	P^a(x)\Bigl),
\end{align}
with
\begin{align}
	P^a(x)=\bar q (x) \gamma_5 T^a q(x), \qquad \tilde{\partial}_\mu =
	\frac{1}{2}(\partial_\mu +\partial_\mu^*),
\end{align}
and the renormalization for the octet current is 
\begin{align}
	A^{R,8,I}_\mu &= Z_A\Bigl[\bigl(1+ 3\overline{b}_A \, a m_q^{\rm av}   +\frac{b_A}{3} \,
	a (m_{l} + 2 m_{s})\bigr) A^{8,I}_\mu \nonumber\\
	&+ 2\left(\frac{1}{3} b_A+ f_A\right)
	 a (m_{l} -m_{s}) A^{0,I}_\mu\Bigr],
\end{align}
with $m_q^{\rm av} = (2 m_{l}+m_{s})/3$ and
where we use the improvement coefficients for $c_A$, $b_A$ from
\cite{Bulava:2015bxa,Bali:2021qem},  and the  renormalization constant of the axial-vector current  $Z_A$ from
\cite{DallaBrida:2018tpn}, respectively. We expect the coefficient
$\overline{b}_A$ to be small, as it is a genuine sea-quark effect, and since a
non-perturbative determination is not available we will neglect
it.
For more details on the calculation of the connected and disconnected pieces we
refer to \cite{Djukanovic:2022wru,Agadjanov:2023efe,Capitani:2017qpc,Ce:2022eix}.

\section{Analysis strategy\label{sec:analysis}}
The analysis strategy follows closely the one of~\citep{Djukanovic:2022wru,Barone:2025hhf}. Namely, we make use of the summation method~\cite{Maiani:1987by,Capitani:2012gj} to address the excited-state contamination and perform a linear fit to the summed expression
\begin{align}
\label{eq:summation}
 S(t_s, \mathbf{q}) 
	& \stackrel{\hphantom{t_s\gg 1}}{=}  a  \sum_{t=a}^{t_s-a}
	G_A^{\mathrm{eff}}(t, t_s,\mathbf{q}) \\ \notag
  & \stackrel{t_s\gg 1}{=} b_0(Q^2) + t_s G_A(Q^2) 
  \, +\mathcal{O}(t_s e^{-\Delta t_s}) \, ,
\end{align}
where $G_A^{\mathrm{eff}}$ is the effective form factor extracted from the
ratios Eq.~(\ref{eq:ratio}), either applying suitable projections or solving
a system of equations explicitly.
We fit simultaneously for the ranges in $Q^2$ and $t_s$ using two parameterizations for the form factor, i.e.  $z$-expansion
\begin{align}
\label{eq:GA-zexp}
G^z_A(Q^2) = \sum_{k=0}^{n} a_k z^{k}(Q^2) \, ,
\end{align}
and a dipole modified with a $z$-expansion
\begin{align}
\label{eq:GA-dipzexp}
G^{dz}_A(Q^2) = \frac{1}{(1 + Q^2/M^2)^2} \left( \sum_{k=0}^{n-1} a_k z^{k}(Q^2) \right) \, , 
\end{align}
where $M$ is the axial mass and
\begin{align}
\quad z(Q^2) = \frac{\sqrt{t_{\rm cut}+Q^2}-\sqrt{t_{\rm cut}}}{\sqrt{t_{\rm cut}+Q^2}+\sqrt{t_{\rm cut}}} \, .
\end{align}
More precisely, we fix $n=2$ for both ans\"atze, substitute them in~\eqref{eq:summation} and fit for all source-sink separations $t_s$ above all possible choices of minimum $t_{s,\rm min}$ and up to $Q^{2}=0.7\,\text{GeV}^2$. We set  $t_{\rm cut} = (4M_{\pi})^2$ using the physical pion mass $M_\pi=134.977\,\text{MeV}$ across all the ensembles in order to simplify the chiral extrapolation. This choice also defines the physical pion mass in our work.
We introduce a broad Gaussian prior for the axial mass $M^{\rm prior}[\text{GeV}] \sim \mathcal{N}(1.285, 0.2)$ in order to center it on the $f_1(1285)$
resonance~\cite{ParticleDataGroup:2024cfk}. The presence of the prior does not affect the final shape of the effective form factors,
but it helps stabilize the fitting procedure by including physical information about the process.

\begin{figure}
\centering
\includegraphics[scale=0.25]{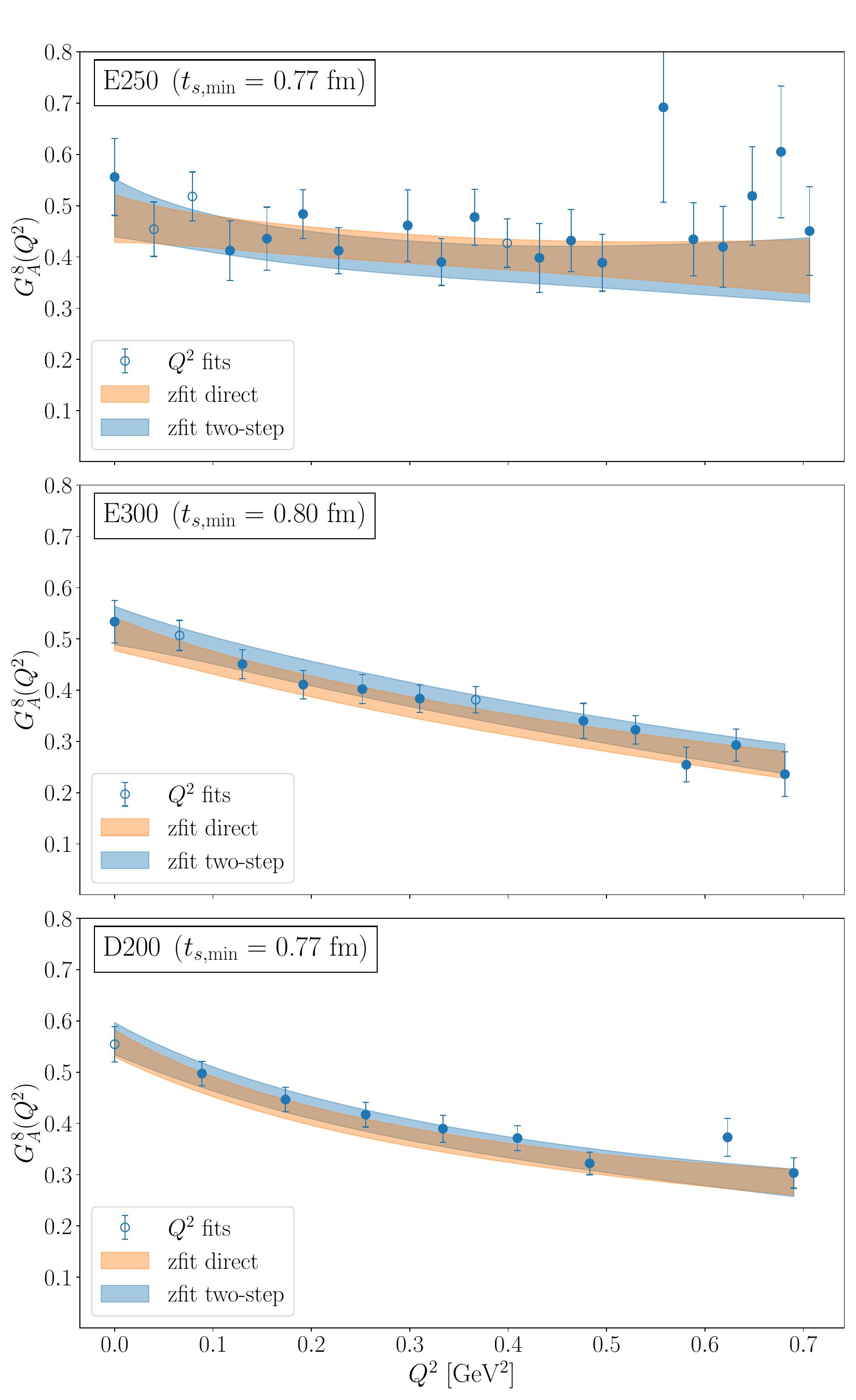}
\caption{Comparison of the direct $z$-fit and the two-step procedure for the ensembles
E250  at $t_{s,\mathrm{min}}=0.77\,\text{fm}$ (top),
E300  at $t_{s,\mathrm{min}}=0.80\,\text{fm}$ (center),
D200  at $t_{s,\mathrm{min}}=0.77\,\text{fm}$ (bottom).
The empty points indicate a p-value smaller than $5\%$  or larger than $95\%$ for the first step of the two-step procedure, i.e. the linear fit to the
summation expression in~\eqref{eq:summation}.
}
\label{fig:plot_zfit_3ens}
\end{figure}

\begin{figure}[tp]
 \includegraphics[scale=0.25]{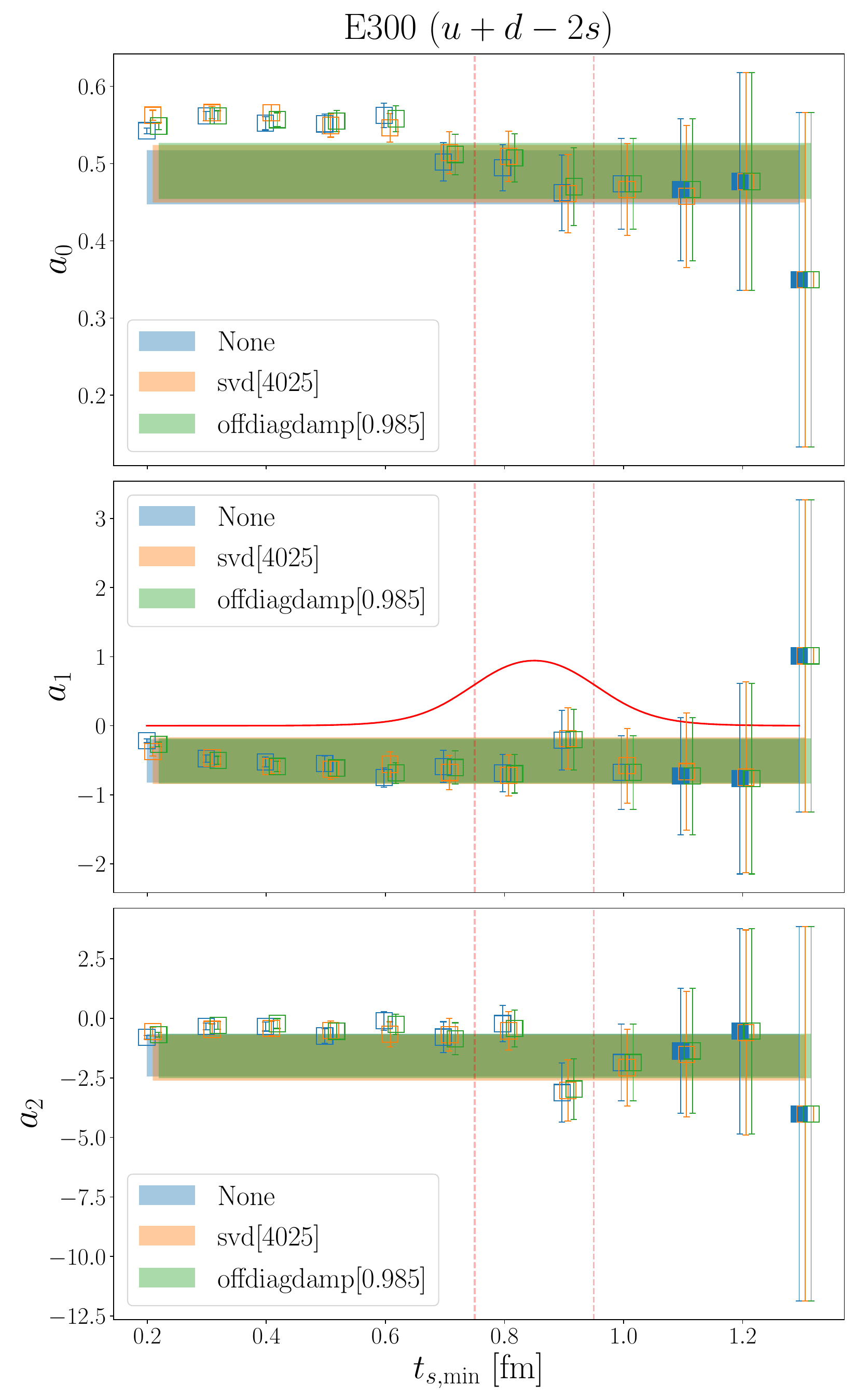}
 \caption{
 Results of the direct fit to the coefficients of the $z$-expansion as a function of the minimum source-sink separation $t_{s,\rm min}$
 for different regularizations of the covariance matrix on the E300 ensemble.
 The numbers in brackets next to ``svd'' and ``offdiagdamp''represent the chosen condition number to perform the SVD cut
 and the off-diagonal damping, respectively. For the unregulated case, labeled as ``None'', filled symbols indicate that the $p$-value is between $0.05$ and $0.95$. The horizonal bands illustrate the final result of the window average on all points. The red curve represents an enlarged version of the window~\eqref{eq:window}, and the vertical lines correspond to $t_w^{\rm low}$
 and $t_w^{\rm up}$.
}
 \label{fig:window_E300}
\end{figure}

The direct parameterization of the form factor with the above expressions has a
stabilizing effect in the fitting strategy. On the other hand, it requires the
estimation of a large covariance matrix, which may become noisy or
ill-conditioned, possibly leading to unstable results due to the most poorly
estimated lowest eigenvalues having the largest influence on $\chi^2$.
We therefore use the singular value decomposition (SVD) to reduce the influence
of these poorly-known and noisy contributions and to
reduce the condition number of the covariance matrix,
requiring it to be the same for all choices of $t_{s,\rm min}$ entering the linear fit to the summation expression in~\eqref{eq:summation} within a given ensemble. 
The results from this approach are compatible with the two-step procedure of first fitting the $t_s$ range at a fixed $Q^2$ and then performing a $z$-expansion on those points.
As an example, we show a comparison for three of our most chiral ensembles in~\figref{fig:plot_zfit_3ens} for a given choice of $t_{s, \mathrm{min}}$.
Furthermore, the results obtained with an SVD cut are compatible with the values obtained by introducing a small damping in the off-diagonal elements, as shown in~\figref{fig:window_E300}
(cf. also with~\cite{Djukanovic:2022wru}). The results of the fits performed with an SVD cut, which are the ones used for our final analysis, are tabulated in the Appendix in~\tabref{tab:coefficients_zfit} ($z$-expansion) and
in~\tabref{tab:coefficients_dipoleZfit} (modified $z$-expansion) ensemble-by-ensemble.
We refer to the Appendix for more details on the fitting strategy. For completeness, we also report the values of $G^8_A(Q^2)$ obtained from the linear fit to the summation expression in~\eqref{eq:summation} at fixed $Q^2$ for the most relevant choices of
$t_{s,\mathrm{min}}$ in~\tabsref{tab:Q2points_H102}{tab:Q2points_E300}.

The choice of $t_{s, \min}$ influences both the contamination by excited states and the statistical error. In order to strike the right balance between the two, we perform a weighted average
over all the results assigning the weights through a window function
\begin{align}
\label{eq:window}
 \frac{1}{N_w}\left[ \tanh\left(\frac{t_{s,{\rm min}}-t_w^{\rm low}}{\Delta t_w} \right) - \tanh\left(\frac{t_{s,{\rm min}}-t_w^{\rm up}}{\Delta t_w} \right) \right] \, ,
\end{align}
where $N_w$ is a normalization factor and 
$t_w^{\rm low} = 0.75 \, \text{fm}$, $t_w^{\rm up} = 0.95\, \text{fm}$, $\Delta t_w = 0.1\, \text{fm}$ on each ensemble to reduce any human bias.
In~\figref{fig:window_E300} we show an example of the procedure on the ensemble E300, which illustrates that the chosen window correctly assigns more weight to the points that start exhibiting a plateau, i.e. the region where the excited states are efficiently suppressed, before entering the large source-sink separation region, where the error increases rapidly. In addition, we show that different regularizations of the covariance matrix (as opposed to the straightforward jackknife estimation, labeled as ``None'')
produce a negligible effect, proving that the fits are stable and reliable regardless of the chosen strategy to estimate the covariance matrix.

\subsection{Physical-point extrapolation\label{sec:ccf}}

\begin{figure*}[p]
 \centering
 \includegraphics[scale=0.23]{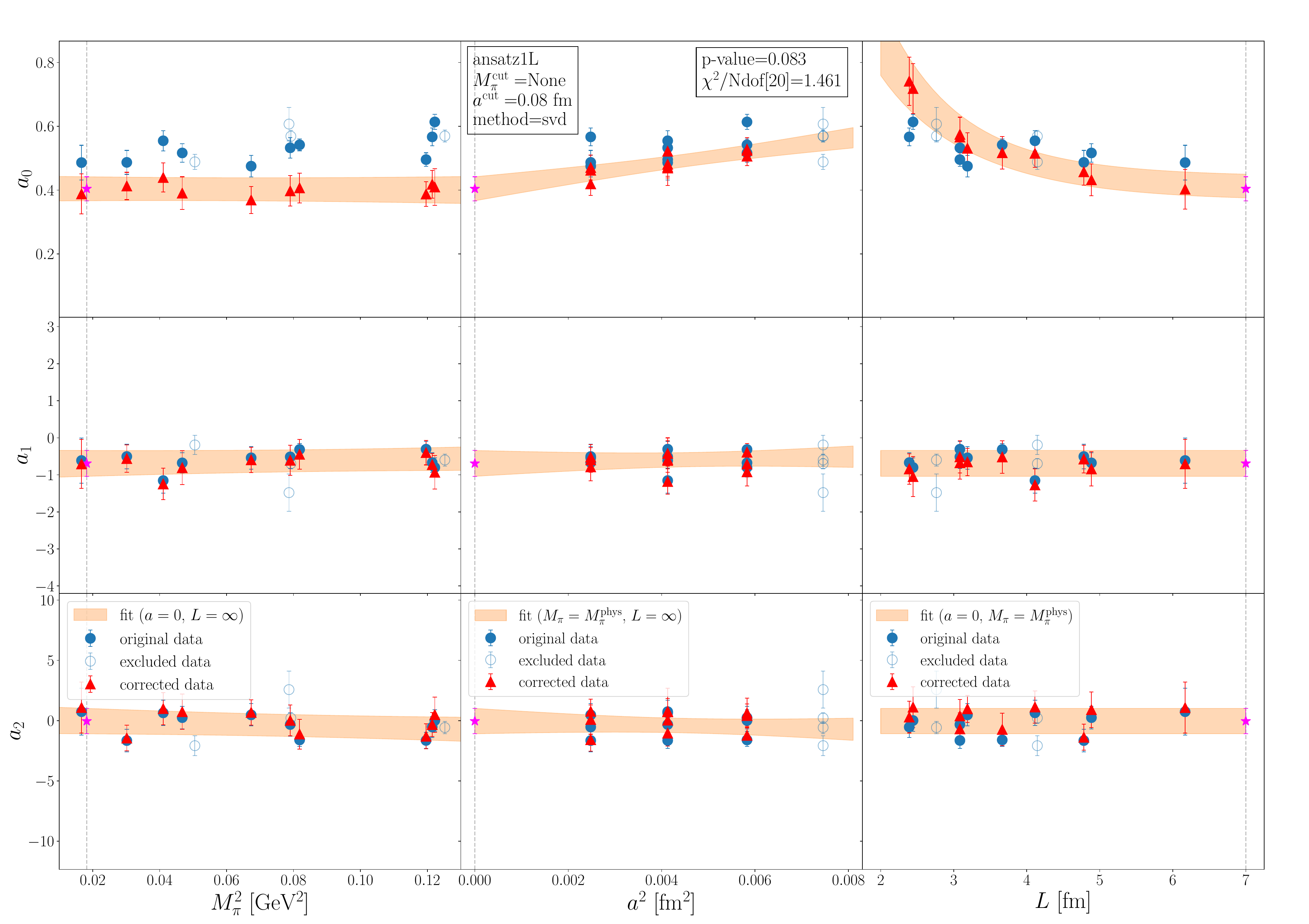}
 \includegraphics[scale=0.23]{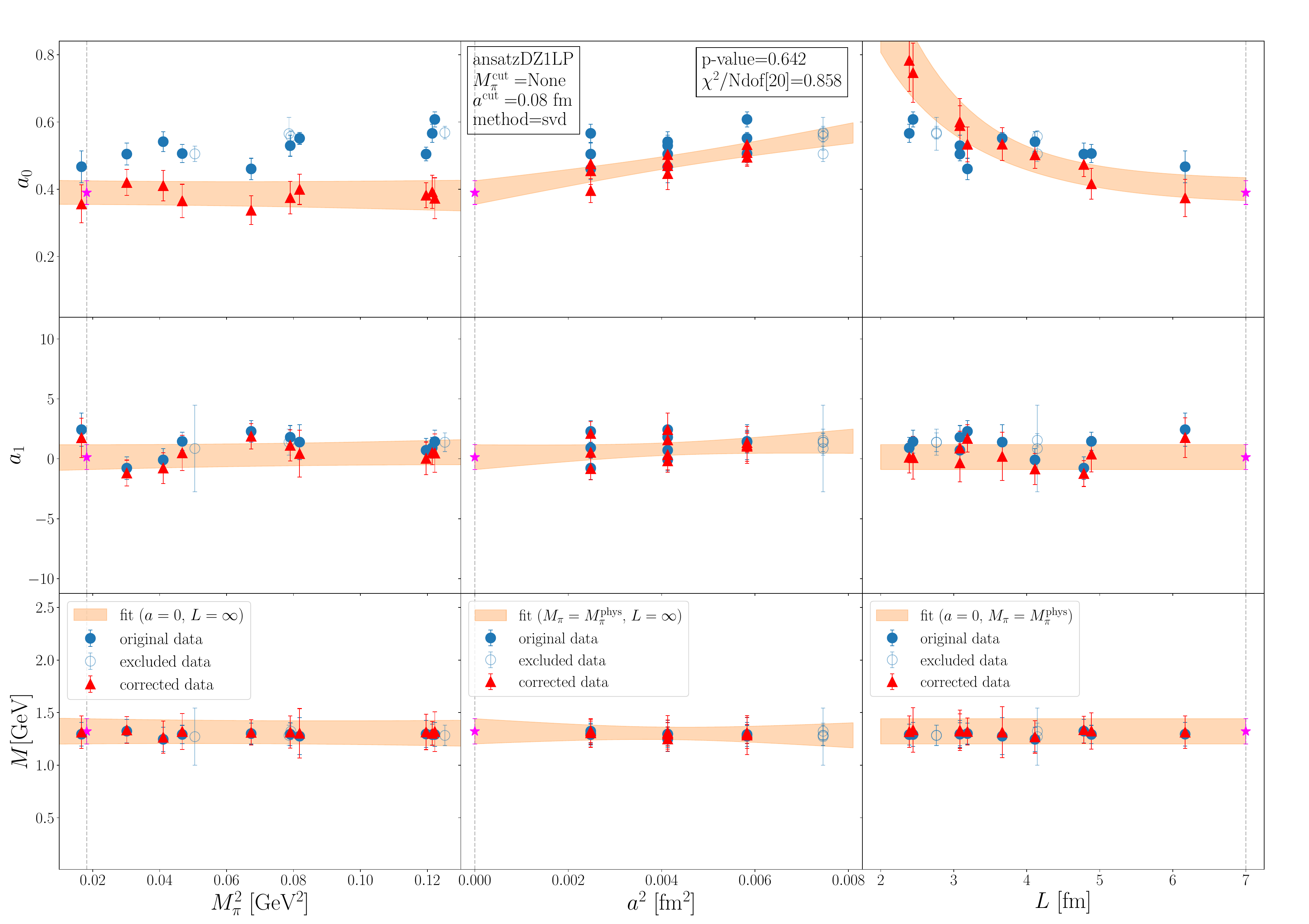}
 \caption{Chiral continuum extrapolation for the $z$-expansion (upper plot) and the modified $z$-expansion (lower plot) ans\"atze 
 with finite-volume term, with no cut on $M_{\pi}$ and without the coarsest lattice spacing. The results are plotted as a function of
 $M^2_{\pi}$ (left), $a$ (center) and $L$ (right). The blue circles indicate the original data, and the filled ones are the ones used after applying the cuts; the red triangles
 represent the corrected data with respect to the fit ansatz, as indicated in the legend for the orange bands.}
 \label{fig:cc_extr}
\end{figure*}

We obtain the coefficients for both fit ans\"atze for $G_A$ through the procedure described above and then
extrapolate the coefficients of these $z$-expansions to the chiral and continuum limit using a simple ansatz linear in $M^2_\pi$ and $a^2$ for all the coefficients $a_i$ (and axial mass $M$), i.e.
\begin{align}
 a_i = d_{i,0} + d_{i,\pi} M^2_\pi +  d_{i,a} a^2 \, .
\end{align}
We include the term~\cite{Beane:2004rf}
\begin{align}
 \frac{M_\pi^2}{\sqrt{M_\pi L}}e^{-M_\pi L}  
\end{align}
in the axial charge $a_0$ to account for possible finite-volume effects.\footnote{We do not include any higher order terms~\cite{Hall:2025ytt} as we are not sensitive to them with our current statistics.}
We perform multiple fits with cuts in the pion mass, $M^{\rm cut}_{\pi}[\text{MeV}]=\{300, 285, 265\}$, and by removing the coarsest lattice spacing, preserving the correlations
among the three coefficients on each ensemble.

This ansatz appears to be sufficient to describe the data well thanks to their flat behavior in the variables $M^2_\pi$, $a^2$ and the
lattice spatial extent $L$. We show an example of the chiral-continuum extrapolation in~\figref{fig:cc_extr} for the $z$-expansion and modified $z$-expansion ans\"atze.

\subsection{Model average}

We perform a weighted average~\cite{Jay:2020jkz} inspired by the Akaike Information Criterion (AIC)~\cite{Akaike}
over the resulting $a_i$
in the version proposed in~\cite{Borsanyi:2020mff}.
In particular, we assign to each fit $k$ the weight
\begin{align}
 w^{\rm AIC}_k \propto 
 e^{-\frac{1}{2}(\chi^2_k +2n_{{\rm par}, k} -n_{{\rm data}, k} )}  \, ,
\end{align}
with $n_{{\rm par}, k}$ being the number of parameters and $n_{{\rm data}, k}$ the number of data points entering the fit.
With these we build the
weighted joint distribution
\begin{align}
\label{eq:weighted_multiv_distro}
 D(\bm{x}) = \sum_{k} w^{\rm AIC}_k
 \mathcal{M}(\bm{x}; \langle \bm{x}^{(k)}\rangle, \cov^{(k)})
\end{align}
for each of the coefficients, where $\langle \bm{x}^{(k)} \rangle$ indicates the mean value of the coefficients $\bm{x}$ in the $k$-th fit 
and $\cov^{(k)}$ is the corresponding covariance matrix of the multivariate normal distribution $\mathcal{M}$. In this case, depending on the fit ansatz for $G_A(Q^2)$,
we have either $\bm{x} = (a_0, a_1, a_2)$ or $\bm{x} = (M, a_0, a_1)$. The final results for the coefficients are obtained from the standard definition
of mean and covariance: labeling two generic coefficients as $y$ and $z$, we have
\begin{align}
 \langle y \rangle = \int \dd y \dd z\, y D(y, z) = \sum_k w^{\rm AIC}_k \langle y^{(k)} \rangle
\end{align}
for the mean value $\langle y\rangle$, and
\begin{align}
 \text{Cov}[y, z]  = \langle yz \rangle - \langle y \rangle \langle z \rangle
\end{align}
for the covariance of each pair of variables $y$ and $z$.
We find that this procedure is essentially equivalent to the one of~\cite{Borsanyi:2020mff}, with
the advantage that it includes both statistical and systematic error automatically, while at the same time preserving the correlations
among the coefficients of the $G_A(Q^2)$ parameterization, as discussed in the Appendix.

\section{Results\label{sec:results}}

\begin{figure}
 \includegraphics[scale=0.3]{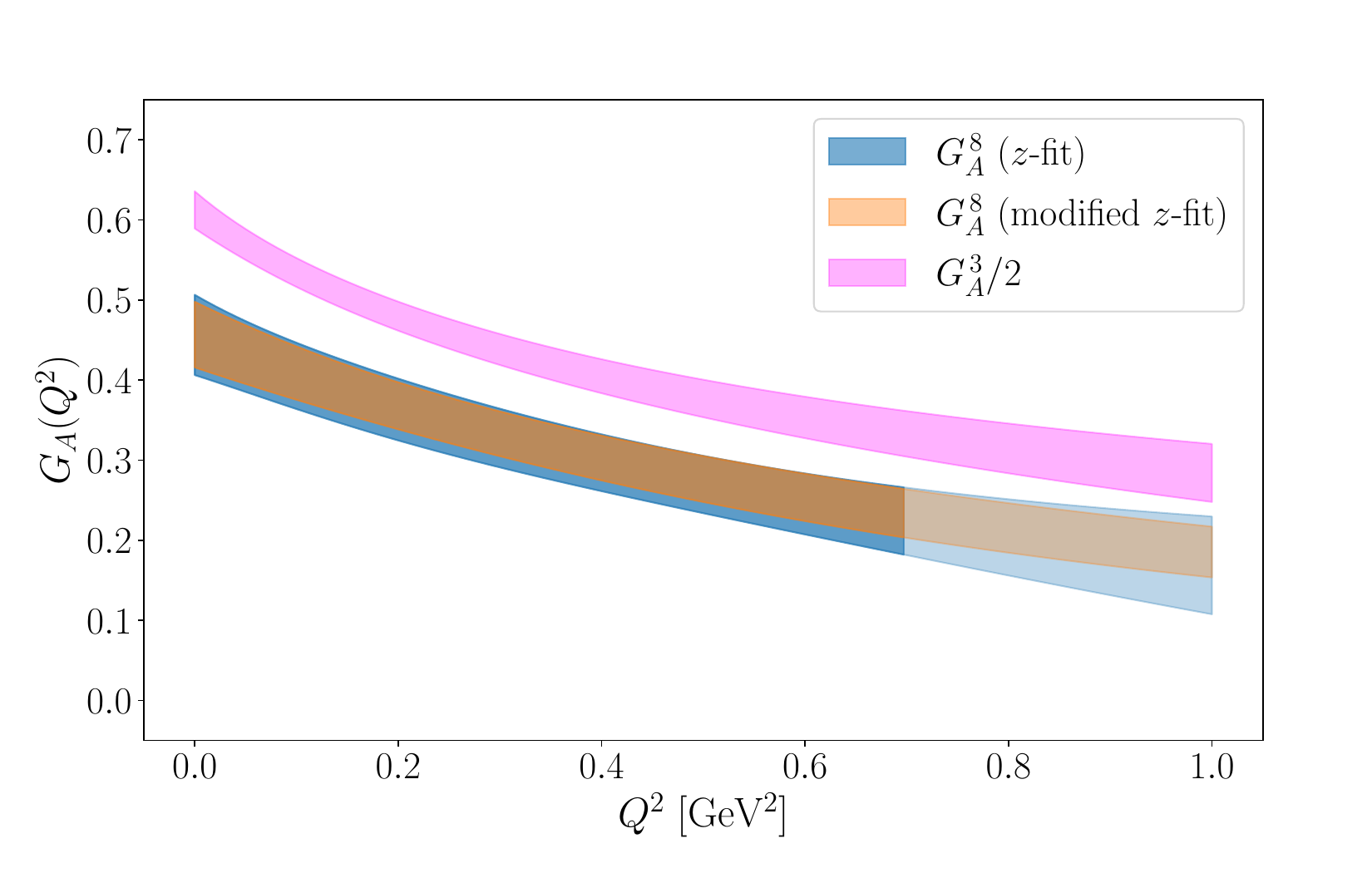}
 \caption{Final form factor for both ans\"atze. The data are used up to $Q^2=0.7\, \text{GeV}^2$, and the lighter-shaded bands represent extrapolations based
   on the respective parameterizations.}
 \label{fig:GA_final}
\end{figure}

Our results for the coefficients of the $z$-expansion of
the nucleon axial form factor in the continuum and at the physical pion mass are
\begin{align}
\begin{split}
a_0 &= +0.456 \pm 0.038 \, \text{(stat)} \pm 0.035 \,\text{(sys)} \,, \\
a_1 &= -0.664 \pm 0.349 \, \text{(stat)} \pm 0.141 \,\text{(sys)} \,, \\
a_2 &= -0.405 \pm 1.067 \, \text{(stat)} \pm 0.531 \,\text{(sys)} \,,
\end{split}
\end{align}
with correlation matrix
\begin{align}
C^{z} = 
\begin{pmatrix}
1.00000 & -0.52277 & 0.23098 \\
-0.52277 & 1.00000 & -0.88683 \\ 
0.23098 & -0.88683 & 1.00000   
\end{pmatrix} \, .
\end{align}
For the modified $z$-expansion we have
\begin{align}
\begin{split}
a_0 &= 0.457 \pm 0.032 \, \text{(stat)} \pm 0.024 \,\text{(sys)} \,, \\
a_1 &= 0.052 \pm 0.909 \, \text{(stat)} \pm 0.182 \,\text{(sys)} \,, \\
M[\text{GeV}] &= 1.312 \pm 0.104 \, \text{(stat)} \pm 0.014 \,\text{(sys)} \,,
\end{split}
\end{align}
with correlation matrix
\begin{align}
C^{dz} = 
\begin{pmatrix}
1.00000 & -0.29470 & -0.03725 \\
-0.29470 & 1.00000 & -0.64760 \\
-0.03725 & -0.64760 & 1.00000 
\end{pmatrix} \, .
\end{align}
The results for both ans\"atze are plotted in~\figref{fig:GA_final} and extrapolated up to $Q^2=1\,\text{GeV}^2$ (as indicated by the shaded areas). We also include
the band for the isovector form factor $G_A^{3}(Q^2)$ from~\cite{Djukanovic:2022wru} divided by a factor of 2 to compare the fall-off between the two channels.

For the octet axial charge we quote
\begin{align}
 g_A^{8}=0.457(40) \,,
\end{align}
which is our most precise result based on the modified $z$-expansion, well consistent with the determination from
the $z$-expansion ansatz $g_A^{8,z}=0.456(51)$. The values are in agreement 
with the prediction $g_A^8=0.46(5)$ obtained using the Cloudy Bag model~\cite{Bass:2009ed}, as well as with the first estimate
from the ETMC Collaboration~\cite{Alexandrou:2020okk} and their most recent result $g_A^8=0.490(20)$~\cite{Alexandrou:2024ozj}.
In contrast, the value of $g_A^8=0.588(33)$~\cite{Deur:2018roz,Close:1993mv}
extracted from hyperon decays assuming SU(3) flavor symmetry is disfavored at the $2.5\sigma$ level.

\section{Conclusions\label{sec:conclusions}}
This work provides the first physical predictions from lattice QCD of the octet axial form factor $G_A^{\,8}(Q^2)$ in a large $Q^2$ range.
We made use of the strategies developed on a companion set of data for the isovector counterpart $u-d$~\cite{Djukanovic:2022wru}, extending some of the key features.
In particular, we parameterized the form factor with either the $z$-expansion or a dipole multiplied by a $z$-expansion, substituting this parameterization directly into the summation expression to perform a simultaneous fit in $Q^2$ and $t_s$.
We reported the final coefficients obtained through a linear ansatz in $M_{\pi}^2$ and $a^2$ for the chiral, continuum and infinite-volume extrapolation, followed by a model average inspired by the AIC.
The results include both systematic and statistical errors, and the value obtained for the axial charge $g_A^{8}$ is compatible with previous results.

This work provides a further step towards the flavor decomposition of the axial form factor,
the completion of which requires a similar analysis for the singlet $u+d+s$ case. 
The strange-quark contribution is of particular interest, as it provides insight into the composition of the proton spin and serves as theoretical input
for neutrino experiments.
An important task for the near future is therefore the computation of the appropriate renormalization factors for the singlet form factor on all ensembles.

\section*{Acknowledgments}
This work was supported
by the Deutsche
  Forschungsgemeinschaft (DFG) through the Collaborative Research
  Center  1660 ``Hadrons and Nuclei as Discovery Tools'',
  under Grant HI~2048/1-3 (Project No.\ 399400745) and in the Cluster
  of Excellence “Precision Physics, Fundamental Interactions and
  Structure of Matter” (PRISMA+ EXC 2118/1) funded by the DFG within
  the German Excellence strategy (Project ID 39083149).
  Calculations for this project were partly performed on the HPC
  clusters ``Clover'' and ``HIMster2'' at the Helmholtz Institute Mainz,
  and ``Mogon 2'' at Johannes Gutenberg-Universit\"at Mainz.
  The authors gratefully acknowledge the Gauss Centre for Supercomputing e.V. (www.gauss-centre.eu) 
  for funding this project by providing computing time on the GCS Supercomputer systems JUQUEEN and JUWELS at J\"ulich Supercomputing Centre (JSC) 
  via Grants HMZ21, HMZ23 and HMZ36 (the latter through the John von Neumann Institute for Computing (NIC)), as well as on the GCS Supercomputer HAZELHEN at
   H\"ochstleistungsrechenzentrum Stuttgart (www.hlrs.de) under Project GCS-HQCD.
  
Our programs use the QDP++ library~\cite{Edwards:2004sx} and deflated SAP+GCR
solver from the openQCD package~\cite{Luscher:2012av}, while the contractions
have been explicitly checked using~\cite{Djukanovic:2016spv}. We are grateful to
our colleagues in the CLS initiative for sharing the gauge field configurations
on which this work is based.

\cleardoublepage
\section*{\label{sec:appendix}Appendix}

\subsection{Method}
\label{sec:app-method}

\begin{figure}[th!]
 \cleardoublepage
 \centering
 \includegraphics[scale=0.3]{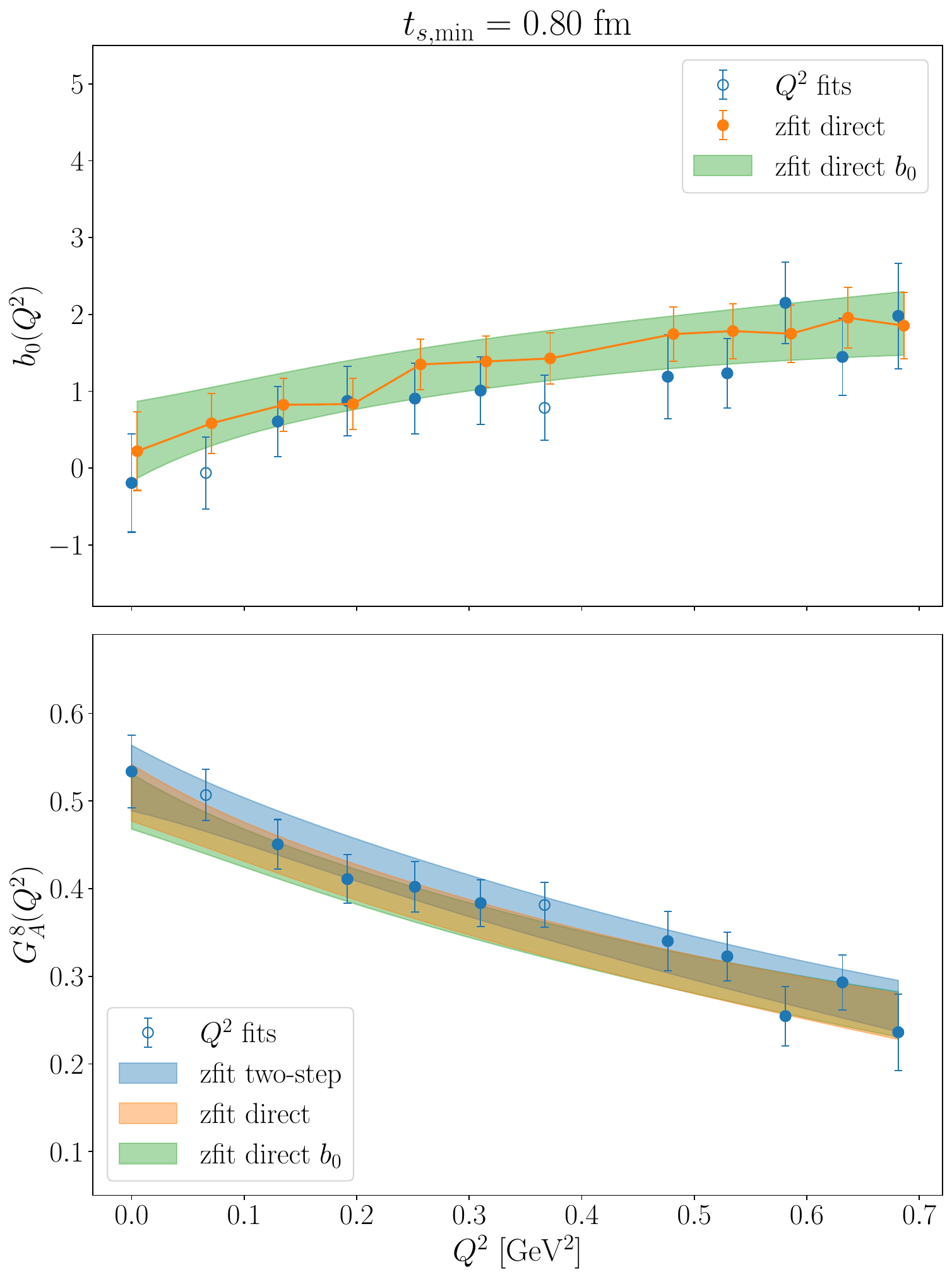}
 \renewcommand{\thefigure}{8}
 \caption[19]{Comparison of different $z$-fit procedures for the ansatz~\eqref{eq:summation} for the extraction of the form factor $G_A^{8}(Q^2)$ (lower plot) on E300. The blue points refer to the two-step procedure, whereas the orange points refer to the direct approach. The green bands complement the latter considering the case where $b_0(Q^2)$ (upper plot) is also parameterized by a $z$-expansion of order $n_b=2$.
 The empty points indicate a $p$-value larger than $95\%$ for the first step of the two-step procedure, i.e. the linear fit to the summation expression in~\eqref{eq:summation}, 
 highlighting another advantage of using the direct approach.}
 \label{fig:zfit-comparison}
\end{figure}

As described in the main text, our fits rely on a (modified) $z$-expansion for the form factor directly incorporated
into the fit to~\eqref{eq:summation} performed over a simultaneous range of both $Q^2$ and $t_{s}$. This has already been proven to give
stable results compatible with the standard two-step approach of first fitting the $t_{s}$ range at a fixed $Q^2$ and then performing the $z$-expansion
on those points~\cite{Djukanovic:2022wru}. Here we further consider the possibility of incorporating a $z$-expansion for the parameter $b_0$ in~\eqref{eq:summation},
namely $b_0(Q^2) = \sum_{k=0}^{n_b} d_k z^{k}(Q^2)$ at order $n_b=2$, to further explore the stability of the fits. An example is shown in~\figref{fig:zfit-comparison}, where
the two-step procedure (blue points) is compared with our direct approach, with or without a $z$-expansion parameterization for the term $b_0$. All the approaches are shown to give compatible results and we do not observe any significant deviation among them on any ensembles. 

\begin{figure}[t]
\centering
\includegraphics[scale=0.28]{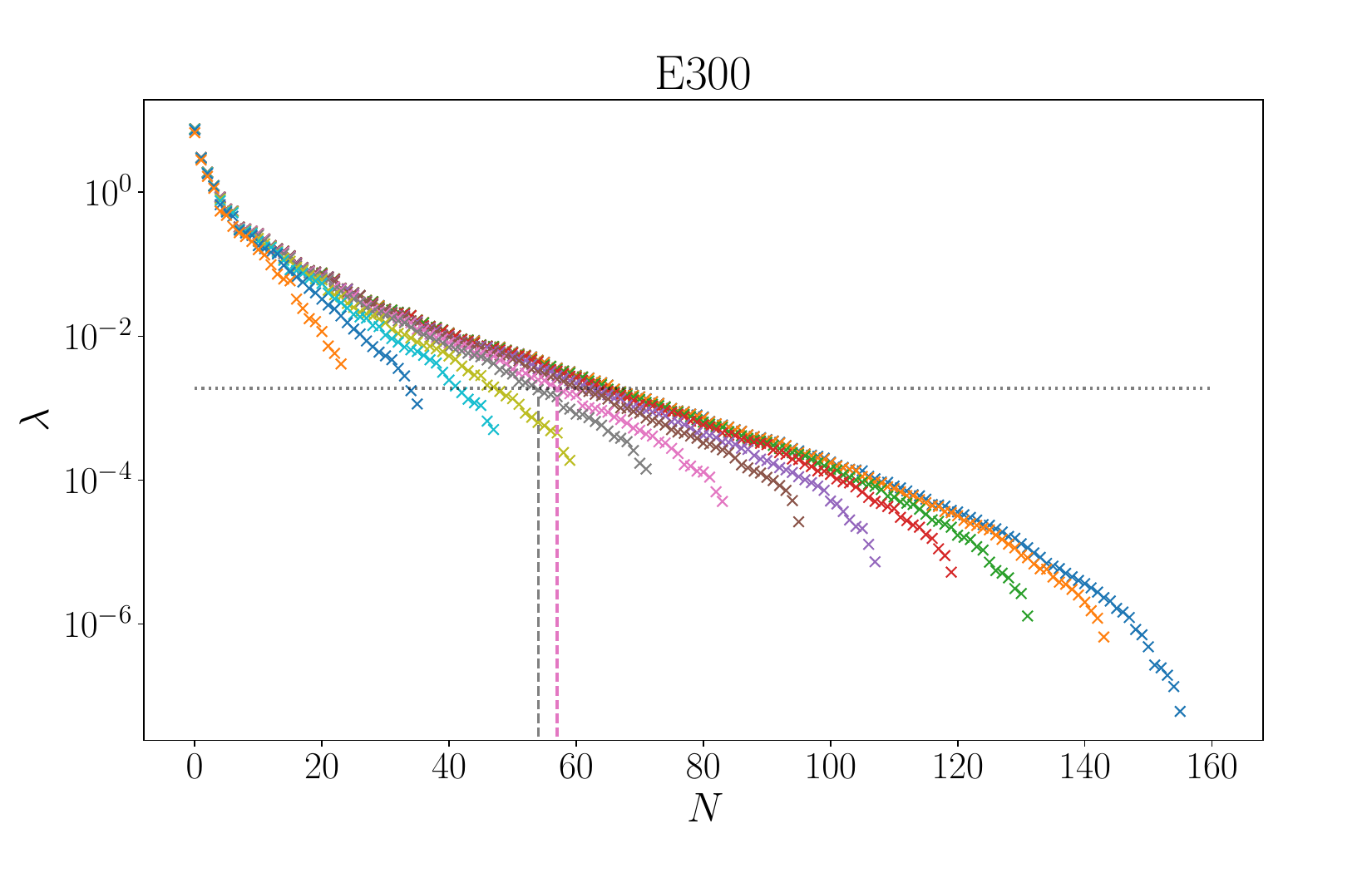}
\renewcommand{\thefigure}{9}
\caption{Example of the scan performed to determine the SVD cut on E300.
The $x$ axis labels the index of the singular values for each of the choices of
$t_{s,\rm min}/a\in \{4, 6, ..., 26\}$, indicated with different colors, while the $y$ axis indicates the singular values of the matrix.
Thus, the last dot of each color indicates the size of the corresponding covariance matrix.
The horizontal line corresponds to our choice for the cut, applied to all values of $t_{s,\rm min}$ in the analysis.
The vertical lines indicate the point where the cut is applied on the two $t_{s,\rm min}$ in the center of the window average, i.e. 
$t_w^{\rm low} \leq t_{s,\rm min} \leq t_w^{\rm up}$.
}
\label{fig:svd_scan}
\end{figure}

We also explored more options to regulate a large covariance matrix, introducing either an off-diagonal damping or an SVD cut.
The latter have been determined by removing the smallest singular values $\lambda$
iteratively (and setting all values $\lambda<\lambda_{\rm min}$ to $\lambda=\lambda_{\rm min}$)
until the fit becomes stable against poorly-estimated fluctuations from the
lowest eigenmodes, as shown in~\figref{fig:svd_scan}. 
In order to estimate when this aim has been realized, we consider the reduced
$\chi^2$ of the fit as an indicator of the stability of the fit, although we
note that this loses its statistical meaning in the presence of SVD cuts.
Operationally, we define a $p$-value  derived from the reduced $\chi^2$ to be acceptable if it lies between $0.05$ and $0.95$,
irrespective of whether 
the covariance matrix has been regulated or not.
We find that the choice of the cut has little impact on the final outcome. The
results are in agreement both with the naive estimate of the covariance matrix
from the jackknife analysis (which in some cases leads to an unacceptable
$\chi^2$) and the introduction of a small off-diagonal damping
(see~\figref{fig:window_E300}). This agreement indicates that the results
obtained are not significantly affected by the poorly estimated lowest
eigenmodes of the covariance matrix, and that we are justified in employing SVD
cuts to suppress the impact of these modes without significantly disturbing the final result.

\subsection{Model average}
\label{sec:app-model-average}

In this work we perform a simple AIC-inspired weighted average over all the variations in the chiral-continuum extrapolation,
which gives compatible a result with the strategy of~\cite{Borsanyi:2020mff}. The latter
involves the construction of a cumulative distribution function (CDF)
\begin{align}
\label{eq:cumulative}
 P(x) = \int_{-\infty}^{x} \dd x'\, \sum_{k} w^{\rm AIC}_k
 \mathcal{N}(x'; \langle x^{(k)}\rangle,  \sigma_{x^{(k)}})
\end{align}
for every coefficient $x^{(k)}$ of each $k$-th fit. The final mean $\langle x\rangle$ is determined from the $50$th percentile, i.e. $\langle x\rangle\equiv x|_{50}$ such that
$P(x|_{50})=0.5$, and the error is obtained similarly from the $16$th and $84$th percentiles, i.e. $\sigma_x = (x|_{84}-x|_{16})/2$. The latter already contains
both systematic and statistical error, which can be disentangled repeating the procedure by uniformly inflating the variances in~\eqref{eq:cumulative}
$\sigma_{x^{(k)}} \rightarrow \lambda\sigma_{x^{(k)}}$, $\lambda\neq 1$,
to assess 
the magnitude of the systematic effects, which are independent from $\lambda$.
To obtain the correlations among coefficients, labeled generically as $x$ and $y$, the procedure must be repeated by building a 
new CDF for the product $z=xy$.

\begin{figure}[t]
\centering
\includegraphics[scale=0.3]{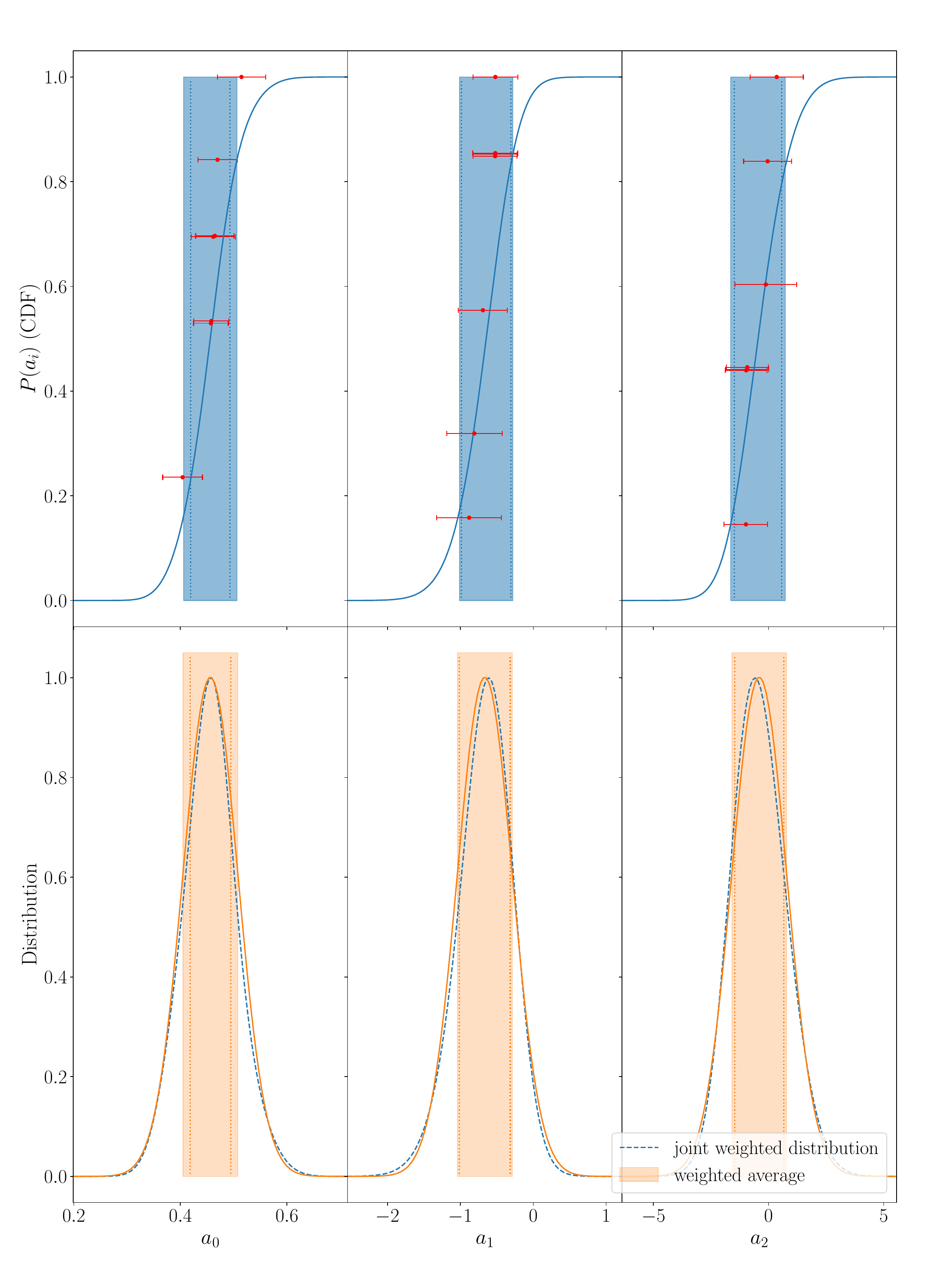}
\caption{Final result obtained after model average with AIC (blue bands, upper plots) or weighted average (orange bands, lower plots) of the three coefficients $a_0$, $a_1$, $a_2$ of the $z$-expansion. The vertical dotted lines indicate the statistical error for both averages.}
\label{fig:AIC}
\end{figure}

However, the simpler mean and error estimation of a weighted sum of the multivariate distributions of the coefficients obtained from each fit
gives the same results preserving by default the correlations and allowing a simple disentangling of systematic and statistical errors.
We start from the definition
\begin{align}
\label{eq:cov}
 \text{Cov}[x, y]  = \langle xy \rangle - \langle x \rangle \langle y \rangle
\end{align}
with
\begin{align}
 \label{eq:mean}
 \langle x \rangle &= \int \dd x \dd y\, x D(x, y) = \sum_k w^{\rm AIC}_k \langle x^{(k)}\rangle \, ,\\
 \langle xy \rangle &= \int \dd x \dd y\, xy D(x, y) = \sum_k w^{\rm AIC}_k \langle x^{(k)} y^{(k)}\rangle  \, ,
\end{align}
where the distribution $D(x,y)$ has been defined in~\eqref{eq:weighted_multiv_distro}.
It can be shown explicitly that
\begin{align}
 \text{Cov}[x, y] = \text{Cov}^{\rm sys}[x, y] + \text{Cov}^{\rm stat}[x, y] \, ,
\end{align}
where
\begin{align}
 \text{Cov}^{\rm stat}[x, y] 
 & = \sum_k w^{\rm AIC}_k \text{Cov}[x^{(k)}, y^{(k)}] \\ \notag
 & = \sum_k w^{\rm AIC}_k \left[  \langle x^{(k)} y^{(k)} \rangle - \langle x^{(k)} \rangle \langle  y^{(k)} \rangle  \right] \, ,
\end{align}
and
\begin{align}
 \text{Cov}^{\rm sys}[x, y] 
  & = \sum_k w^{\rm AIC}_k \left[ \langle x^{(k)} \rangle -\langle x\rangle  \right]  \left[ \langle y^{(k)} \rangle -\langle y\rangle  \right]  \\ \notag
  & = \sum_k w^{\rm AIC}_k \langle  x^{(k)} \rangle \langle  y^{(k)}   \rangle - \langle x \rangle \langle y \rangle  \, .
\end{align}
We show in~\figref{fig:AIC} the comparison between the two approaches to justify our choice. The upper plot corresponds to the AIC average performed on the fit results (red points) 
through the use of the CDF (blue curve), where the vertical blue bands indicate the final result in terms of an
interval spanning between the $16$th and $84$th percentiles of the distribution in~\eqref{eq:cumulative}.
The lower plot compares the full distribution used to build the CDF (dashed blue curve) to the one obtained with a weighted average of the fit results (solid orange curve).
The vertical dotted lines indicate the statistical error for both cases.
In other words, the orange bands in the lower plot
are built using the mean in~\eqref{eq:mean} and the covariance (variance) in~\eqref{eq:cov}
and assuming the final result follows a Gaussian distribution.
We see that the two approaches yield no significant difference, either in the total error nor the decomposition into statistical and systematic uncertainties.
This relies on the fact that the average of our results behaves like a Gaussian.
Note that this may not be the case in general, e.g. where more fit ans\"atze are used for the chiral-continuum extrapolation,
and may not be applicable in all situations.

\cleardoublepage
\onecolumngrid

\cleardoublepage

\subsection{Form factor data}

We report the numerical results of the form factor extraction on each ensemble as described in~\secref{sec:analysis}. 
In~\tabsref{tab:coefficients_zfit}{tab:coefficients_dipoleZfit} we collect the results of
the simultaneous fits in $Q^2$ and $t_s$ for the $z$-expansion and modified $z$-expansion
ans\"atze  after window average, respectively.
For completeness, in~\tabsref{tab:Q2points_H102}{tab:Q2points_E300} we report the results of $G^{8}_A(Q^2)$ obtained from the linear fit to the summation expression
in~\eqref{eq:summation} at fixed $Q^2$
for $t_w^{\mathrm{low}} \leq t_{s,\mathrm{min}} \leq t_w^{\mathrm{up}}$, with $t_w^{\mathrm{low}}=0.75\,\text{fm}$
and $t_w^{\mathrm{up}}=0.95\,\text{fm}$.

\begin{table}[!hb]
 \centering
\begin{tabular}{lcccccc}
\hline\hline
 & $a_0$ & $a_1$ & $a_2$ & $\rho_{a_0,a_1}$ & $\rho_{a_0,a_2}$ & $\rho_{a_1,a_2}$ \\
\hline\hline
 H102 & $0.569(19)$ & $-0.60(16)$ & $-0.57(52)$ & $-0.55804$ & $0.25473$ & $-0.88293$  \\
 H105 & $0.607(53)$ & $-1.48(51)$ & $2.6(1.5)$ & $-0.68792$ & $0.43632$ & $-0.91044$  \\
 C101 & $0.488(24)$ & $-0.19(25)$ & $-2.08(83)$ & $-0.60011$ & $0.37279$ & $-0.88766$  \\
 N101 & $0.569(18)$ & $-0.70(13)$ & $0.21(43)$ & $-0.60403$ & $0.31946$ & $-0.84955$  \\
 S400 & $0.614(24)$ & $-0.80(26)$ & $0.010(878)$ & $-0.56939$ & $0.31177$ & $-0.91693$  \\
 N451 & $0.542(19)$ & $-0.31(16)$ & $-1.60(52)$ & $-0.61864$ & $0.32671$ & $-0.86284$  \\
 D450 & $0.516(28)$ & $-0.68(27)$ & $0.25(92)$ & $-0.63480$ & $0.35958$ & $-0.87961$  \\
 N203 & $0.496(21)$ & $-0.31(20)$ & $-1.64(65)$ & $-0.58782$ & $0.30535$ & $-0.88447$  \\
 N200 & $0.532(33)$ & $-0.51(32)$ & $-0.3(1.0)$ & $-0.58641$ & $0.31684$ & $-0.89698$  \\
 D200 & $0.555(33)$ & $-1.15(33)$ & $0.6(1.0)$ & $-0.63127$ & $0.38864$ & $-0.89982$  \\
 E250 & $0.486(54)$ & $-0.61(62)$ & $0.7(2.0)$ & $-0.73743$ & $0.52539$ & $-0.92201$  \\
 N302 & $0.567(28)$ & $-0.67(26)$ & $-0.53(86)$ & $-0.56357$ & $0.27283$ & $-0.90206$  \\
 J303 & $0.475(34)$ & $-0.54(31)$ & $0.48(95)$ & $-0.67460$ & $0.40246$ & $-0.88869$  \\
 E300 & $0.487(37)$ & $-0.50(34)$ & $-1.65(96)$ & $-0.69221$ & $0.46788$ & $-0.89354$  \\
\hline\hline
\end{tabular}
  \caption{Results for the coefficients $a_0$, $a_1$, $a_2$ of the $z$-expansion for each ensemble,
  as well as their correlations $\rho$.
  These are obtained from the window average of
  $z$-expansion fits to the sum $S(\vec q,t_s)$ in~\eqref{eq:summation} using different $t_{s,\textrm{min}}$.}
 \label{tab:coefficients_zfit}
\end{table}

\begin{table}[!hb]
 \centering
\begin{tabular}{lcccccc}
\hline\hline
 & $a_0$ & $a_1$ & $M[\text{GeV}]$ & $\rho_{a_0,a_1}$ & $\rho_{a_0,M}$ & $\rho_{a_1,M}$ \\
\hline\hline
 H102 & $0.568(19)$ & $1.38(77)$ & $1.283(95)$ & $-0.18210$ & $-0.04088$ & $-0.87744$  \\
 H105 & $0.565(49)$ & $1.4(1.1)$ & $1.284(98)$ & $-0.42137$ & $-0.20474$ & $-0.39070$  \\
 C101 & $0.505(23)$ & $0.9(3.6)$ & $1.27(27)$ & $-0.34602$ & $0.30670$ & $-0.95599$  \\
 N101 & $0.557(17)$ & $1.55(55)$ & $1.323(79)$ & $-0.13591$ & $-0.10115$ & $-0.79267$  \\
 S400 & $0.608(23)$ & $1.43(95)$ & $1.29(11)$ & $-0.20571$ & $-0.08311$ & $-0.84400$  \\
 N451 & $0.552(18)$ & $1.4(1.5)$ & $1.28(18)$ & $-0.30308$ & $0.18872$ & $-0.96142$  \\
 D450 & $0.506(27)$ & $1.46(76)$ & $1.293(89)$ & $-0.35386$ & $-0.06347$ & $-0.59937$  \\
 N203 & $0.505(21)$ & $0.7(1.0)$ & $1.30(13)$ & $-0.29013$ & $0.08477$ & $-0.89023$  \\
 N200 & $0.530(31)$ & $1.80(99)$ & $1.30(11)$ & $-0.30242$ & $-0.03475$ & $-0.71410$  \\
 D200 & $0.541(31)$ & $-0.09(92)$ & $1.25(12)$ & $-0.23985$ & $-0.18638$ & $-0.60583$  \\
 E250 & $0.467(47)$ & $2.4(1.4)$ & $1.29(11)$ & $-0.50752$ & $-0.03163$ & $-0.49149$  \\
 N302 & $0.566(27)$ & $0.92(85)$ & $1.291(98)$ & $-0.28218$ & $-0.06353$ & $-0.77017$  \\
 J303 & $0.461(31)$ & $2.29(90)$ & $1.301(100)$ & $-0.41322$ & $-0.03005$ & $-0.62716$  \\
 E300 & $0.505(33)$ & $-0.78(97)$ & $1.32(11)$ & $-0.37199$ & $0.04028$ & $-0.64084$  \\
\hline\hline
\end{tabular}
  \caption{Results for the coefficients $a_0$, $a_1$, $M$ of the modified $z$-expansion for each ensemble, with $M$ being the axial mass of the dipole, 
  as well as their correlations $\rho$.
  These are obtained from the window average of
  modified $z$-expansion fits to the sum $S(\vec q,t_s)$ in~\eqref{eq:summation} using different $t_{s,\textrm{min}}$.}
 \label{tab:coefficients_dipoleZfit}
\end{table}

\begin{table}
\begin{tabular}{l|ccc}
\hline\hline
\multicolumn{4}{c}{H102} \\ 
 \hline\hline 
$Q^2$ [GeV$^2$] & \multicolumn{3}{c}{$G^{8}_A(Q^2)$ for $t_{s,\mathrm{min}}$ [fm]} \\ 
  & 0.78 & 0.86 & 0.95 \\
\hline\hline
 0.00 & $0.589(17)$ & $0.558(28)$ & $0.566(31)$  \\
 0.19 & $0.483(15)$ & $0.470(18)$ & $0.480(23)$  \\
 0.37 & $0.425(13)$ & $0.407(20)$ & $0.415(24)$  \\
 0.54 & $0.363(15)$ & $0.371(22)$ & $0.378(27)$  \\
 0.70 & $0.333(20)$ & $0.311(38)$ & $0.318(42)$  \\
\hline\hline
\end{tabular}
\caption{Form factor $G^{8}_A(Q^2)$ extracted from a linear fit of~\eqref{eq:summation} for $t_w^{\mathrm{low}} \leq t_{s,\mathrm{min}} \leq t_w^{\mathrm{up}}$ on H102.}
\label{tab:Q2points_H102}
\end{table}

\begin{table}
\begin{tabular}{l|ccc}
\hline\hline
\multicolumn{4}{c}{H105} \\ 
 \hline\hline 
$Q^2$ [GeV$^2$] & \multicolumn{3}{c}{$G^{8}_A(Q^2)$ for $t_{s,\mathrm{min}}$ [fm]} \\ 
  & 0.78 & 0.86 & 0.95 \\
\hline\hline
 0.00 & $0.533(45)$ & $0.754(79)$ & $0.781(85)$  \\
 0.19 & $0.454(25)$ & $0.497(52)$ & $0.507(57)$  \\
 0.37 & $0.376(27)$ & $0.443(64)$ & $0.444(71)$  \\
 0.54 & $0.354(31)$ & $0.360(64)$ & $0.354(69)$  \\
 0.70 & $0.333(43)$ & $0.47(11)$ & $0.52(12)$  \\
\hline\hline
\end{tabular}
\caption{Form factor $G^{8}_A(Q^2)$ extracted from a linear fit of~\eqref{eq:summation} for $t_w^{\mathrm{low}} \leq t_{s,\mathrm{min}} \leq t_w^{\mathrm{up}}$ on H105.}
\label{tab:Q2points_H105}
\end{table}

\begin{table}
\begin{tabular}{l|ccc}
\hline\hline
\multicolumn{4}{c}{C101} \\ 
 \hline\hline 
$Q^2$ [GeV$^2$] & \multicolumn{3}{c}{$G^{8}_A(Q^2)$ for $t_{s,\mathrm{min}}$ [fm]} \\ 
  & 0.78 & 0.86 & 0.95 \\
\hline\hline
 0.00 & $0.463(32)$ & $0.447(42)$ & $0.450(45)$  \\
 0.09 & $0.454(25)$ & $0.424(39)$ & $0.429(41)$  \\
 0.17 & $0.432(26)$ & $0.358(41)$ & $0.350(43)$  \\
 0.25 & $0.444(25)$ & $0.375(45)$ & $0.370(53)$  \\
 0.33 & $0.378(22)$ & $0.339(42)$ & $0.317(46)$  \\
 0.40 & $0.340(25)$ & $0.289(43)$ & $0.285(44)$  \\
 0.48 & $0.336(25)$ & $0.277(54)$ & $0.220(62)$  \\
 0.62 & $0.280(35)$ & $0.339(82)$ & $0.313(89)$  \\
 0.68 & $0.263(29)$ & $0.235(57)$ & $0.209(61)$  \\
\hline\hline
\end{tabular}
\caption{Form factor $G^{8}_A(Q^2)$ extracted from a linear fit of~\eqref{eq:summation} for $t_w^{\mathrm{low}} \leq t_{s,\mathrm{min}} \leq t_w^{\mathrm{up}}$ on C101.}
\label{tab:Q2points_C101}
\end{table}

\begin{table}
\begin{tabular}{l|ccc}
\hline\hline
\multicolumn{4}{c}{N101} \\ 
 \hline\hline 
$Q^2$ [GeV$^2$] & \multicolumn{3}{c}{$G^{8}_A(Q^2)$ for $t_{s,\mathrm{min}}$ [fm]} \\ 
  & 0.78 & 0.86 & 0.95 \\
\hline\hline
 0.00 & $0.550(21)$ & $0.568(29)$ & $0.617(43)$  \\
 0.09 & $0.504(17)$ & $0.517(24)$ & $0.544(38)$  \\
 0.17 & $0.481(17)$ & $0.483(29)$ & $0.529(42)$  \\
 0.25 & $0.452(14)$ & $0.461(33)$ & $0.535(63)$  \\
 0.33 & $0.437(16)$ & $0.429(27)$ & $0.450(38)$  \\
 0.41 & $0.424(16)$ & $0.423(27)$ & $0.450(37)$  \\
 0.48 & $0.374(16)$ & $0.385(29)$ & $0.408(49)$  \\
 0.62 & $0.372(23)$ & $0.408(41)$ & $0.483(77)$  \\
 0.69 & $0.372(19)$ & $0.401(32)$ & $0.406(53)$  \\
\hline\hline
\end{tabular}
\caption{Form factor $G^{8}_A(Q^2)$ extracted from a linear fit of~\eqref{eq:summation} for $t_w^{\mathrm{low}} \leq t_{s,\mathrm{min}} \leq t_w^{\mathrm{up}}$ on N101.}
\label{tab:Q2points_N101}
\end{table}

\begin{table}
\begin{tabular}{l|cc}
\hline\hline
\multicolumn{3}{c}{S400} \\ 
 \hline\hline 
$Q^2$ [GeV$^2$] & \multicolumn{2}{c}{$G^{8}_A(Q^2)$ for $t_{s,\mathrm{min}}$ [fm]} \\ 
  & 0.76 & 0.92 \\
\hline\hline
 0.00 & $0.596(18)$ & $0.624(36)$  \\
 0.25 & $0.488(12)$ & $0.502(27)$  \\
 0.47 & $0.417(14)$ & $0.437(33)$  \\
 0.68 & $0.383(17)$ & $0.364(43)$  \\
\hline\hline
\end{tabular}
\caption{Form factor $G^{8}_A(Q^2)$ extracted from a linear fit of~\eqref{eq:summation} for $t_w^{\mathrm{low}} \leq t_{s,\mathrm{min}} \leq t_w^{\mathrm{up}}$ on S400.}
\label{tab:Q2points_S400}
\end{table}

\begin{table}
\begin{tabular}{l|cc}
\hline\hline
\multicolumn{3}{c}{N451} \\ 
 \hline\hline 
$Q^2$ [GeV$^2$] & \multicolumn{2}{c}{$G^{8}_A(Q^2)$ for $t_{s,\mathrm{min}}$ [fm]} \\ 
  & 0.76 & 0.92 \\
\hline\hline
 0.00 & $0.543(19)$ & $0.523(26)$  \\
 0.11 & $0.509(15)$ & $0.488(20)$  \\
 0.22 & $0.471(17)$ & $0.448(22)$  \\
 0.32 & $0.439(15)$ & $0.415(20)$  \\
 0.42 & $0.398(19)$ & $0.393(28)$  \\
 0.51 & $0.375(21)$ & $0.352(29)$  \\
 0.60 & $0.346(18)$ & $0.318(24)$  \\
\hline\hline
\end{tabular}
\caption{Form factor $G^{8}_A(Q^2)$ extracted from a linear fit of~\eqref{eq:summation} for $t_w^{\mathrm{low}} \leq t_{s,\mathrm{min}} \leq t_w^{\mathrm{up}}$ on N451.}
\label{tab:Q2points_N451}
\end{table}

\begin{table}
\begin{tabular}{l|ccc}
\hline\hline
\multicolumn{4}{c}{D450} \\ 
 \hline\hline 
$Q^2$ [GeV$^2$] & \multicolumn{3}{c}{$G^{8}_A(Q^2)$ for $t_{s,\mathrm{min}}$ [fm]} \\ 
  & 0.76 & 0.84 & 0.92 \\
\hline\hline
 0.00 & $0.518(42)$ & $0.539(44)$ & $0.587(55)$  \\
 0.06 & $0.486(31)$ & $0.494(33)$ & $0.519(39)$  \\
 0.12 & $0.468(27)$ & $0.477(31)$ & $0.506(40)$  \\
 0.18 & $0.436(29)$ & $0.429(32)$ & $0.436(42)$  \\
 0.24 & $0.452(32)$ & $0.459(38)$ & $0.455(47)$  \\
 0.30 & $0.417(26)$ & $0.419(30)$ & $0.440(42)$  \\
 0.35 & $0.387(23)$ & $0.382(27)$ & $0.369(38)$  \\
\hline\hline
\end{tabular}
\caption{Form factor $G^{8}_A(Q^2)$ extracted from a linear fit of~\eqref{eq:summation} for $t_w^{\mathrm{low}} \leq t_{s,\mathrm{min}} \leq t_w^{\mathrm{up}}$ on D450.}
\label{tab:Q2points_D450}
\end{table}

\begin{table}
\begin{tabular}{l|cc}
\hline\hline
\multicolumn{3}{c}{N203} \\ 
 \hline\hline 
$Q^2$ [GeV$^2$] & \multicolumn{2}{c}{$G^{8}_A(Q^2)$ for $t_{s,\mathrm{min}}$ [fm]} \\ 
  & 0.77 & 0.90 \\
\hline\hline
 0.00 & $0.562(14)$ & $0.481(25)$  \\
 0.16 & $0.487(11)$ & $0.423(19)$  \\
 0.30 & $0.438(10)$ & $0.366(21)$  \\
 0.44 & $0.410(11)$ & $0.374(25)$  \\
 0.58 & $0.346(14)$ & $0.252(31)$  \\
 0.71 & $0.319(14)$ & $0.204(31)$  \\
\hline\hline
\end{tabular}
\caption{Form factor $G^{8}_A(Q^2)$ extracted from a linear fit of~\eqref{eq:summation} for $t_w^{\mathrm{low}} \leq t_{s,\mathrm{min}} \leq t_w^{\mathrm{up}}$ on N203.}
\label{tab:Q2points_N203}
\end{table}

\begin{table}
\begin{tabular}{l|cc}
\hline\hline
\multicolumn{3}{c}{N200} \\ 
 \hline\hline 
$Q^2$ [GeV$^2$] & \multicolumn{2}{c}{$G^{8}_A(Q^2)$ for $t_{s,\mathrm{min}}$ [fm]} \\ 
  & 0.77 & 0.90 \\
\hline\hline
 0.00 & $0.582(22)$ & $0.458(55)$  \\
 0.16 & $0.516(16)$ & $0.487(50)$  \\
 0.30 & $0.467(17)$ & $0.381(44)$  \\
 0.44 & $0.434(18)$ & $0.376(60)$  \\
 0.57 & $0.404(23)$ & $0.425(70)$  \\
 0.70 & $0.388(23)$ & $0.353(68)$  \\
\hline\hline
\end{tabular}
\caption{Form factor $G^{8}_A(Q^2)$ extracted from a linear fit of~\eqref{eq:summation} for $t_w^{\mathrm{low}} \leq t_{s,\mathrm{min}} \leq t_w^{\mathrm{up}}$ on N200.}
\label{tab:Q2points_N200}
\end{table}

\begin{table}
\begin{tabular}{l|cc}
\hline\hline
\multicolumn{3}{c}{D200} \\ 
 \hline\hline 
$Q^2$ [GeV$^2$] & \multicolumn{2}{c}{$G^{8}_A(Q^2)$ for $t_{s,\mathrm{min}}$ [fm]} \\ 
  & 0.77 & 0.90 \\
\hline\hline
 0.00 & $0.554(35)$ & $0.548(53)$  \\
 0.09 & $0.497(24)$ & $0.469(45)$  \\
 0.17 & $0.447(24)$ & $0.405(40)$  \\
 0.26 & $0.417(24)$ & $0.430(41)$  \\
 0.33 & $0.390(26)$ & $0.345(50)$  \\
 0.41 & $0.371(24)$ & $0.325(49)$  \\
 0.48 & $0.322(22)$ & $0.295(47)$  \\
 0.62 & $0.373(37)$ & $0.351(81)$  \\
 0.69 & $0.303(30)$ & $0.227(59)$  \\
\hline\hline
\end{tabular}
\caption{Form factor $G^{8}_A(Q^2)$ extracted from a linear fit of~\eqref{eq:summation} for $t_w^{\mathrm{low}} \leq t_{s,\mathrm{min}} \leq t_w^{\mathrm{up}}$ on D200.}
\label{tab:Q2points_D200}
\end{table}

\begin{table}
\begin{tabular}{l|cc}
\hline\hline
\multicolumn{3}{c}{E250} \\ 
 \hline\hline 
$Q^2$ [GeV$^2$] & \multicolumn{2}{c}{$G^{8}_A(Q^2)$ for $t_{s,\mathrm{min}}$ [fm]} \\ 
  & 0.77 & 0.90 \\
\hline\hline
 0.00 & $0.556(75)$ & $0.65(10)$  \\
 0.04 & $0.454(53)$ & $0.483(71)$  \\
 0.08 & $0.518(48)$ & $0.526(59)$  \\
 0.12 & $0.412(58)$ & $0.349(85)$  \\
 0.15 & $0.436(61)$ & $0.449(74)$  \\
 0.19 & $0.484(48)$ & $0.478(58)$  \\
 0.23 & $0.412(45)$ & $0.388(55)$  \\
 0.30 & $0.462(70)$ & $0.417(85)$  \\
 0.33 & $0.390(46)$ & $0.348(61)$  \\
 0.37 & $0.478(55)$ & $0.436(77)$  \\
 0.40 & $0.427(47)$ & $0.399(66)$  \\
 0.43 & $0.398(67)$ & $0.40(10)$  \\
 0.46 & $0.432(61)$ & $0.427(83)$  \\
 0.50 & $0.389(55)$ & $0.336(79)$  \\
 0.56 & $0.69(18)$ & $0.73(24)$  \\
 0.59 & $0.434(72)$ & $0.33(10)$  \\
 0.62 & $0.420(79)$ & $0.38(10)$  \\
 0.65 & $0.519(96)$ & $0.49(14)$  \\
 0.68 & $0.61(13)$ & $0.66(16)$  \\
 0.71 & $0.450(87)$ & $0.42(11)$  \\
\hline\hline
\end{tabular}
\caption{Form factor $G^{8}_A(Q^2)$ extracted from a linear fit of~\eqref{eq:summation} for $t_w^{\mathrm{low}} \leq t_{s,\mathrm{min}} \leq t_w^{\mathrm{up}}$ on E250.}
\label{tab:Q2points_E250}
\end{table}

\begin{table}
\begin{tabular}{l|cc}
\hline\hline
\multicolumn{3}{c}{N302} \\ 
 \hline\hline 
$Q^2$ [GeV$^2$] & \multicolumn{2}{c}{$G^{8}_A(Q^2)$ for $t_{s,\mathrm{min}}$ [fm]} \\ 
  & 0.80 & 0.90 \\
\hline\hline
 0.00 & $0.580(23)$ & $0.544(41)$  \\
 0.26 & $0.469(16)$ & $0.438(29)$  \\
 0.49 & $0.395(18)$ & $0.324(35)$  \\
 0.71 & $0.351(20)$ & $0.331(44)$  \\
\hline\hline
\end{tabular}
\caption{Form factor $G^{8}_A(Q^2)$ extracted from a linear fit of~\eqref{eq:summation} for $t_w^{\mathrm{low}} \leq t_{s,\mathrm{min}} \leq t_w^{\mathrm{up}}$ on N302.}
\label{tab:Q2points_N302}
\end{table}

\begin{table}
\begin{tabular}{l|cc}
\hline\hline
\multicolumn{3}{c}{J303} \\ 
 \hline\hline 
$Q^2$ [GeV$^2$] & \multicolumn{2}{c}{$G^{8}_A(Q^2)$ for $t_{s,\mathrm{min}}$ [fm]} \\ 
  & 0.80 & 0.90 \\
\hline\hline
 0.00 & $0.480(29)$ & $0.462(52)$  \\
 0.15 & $0.449(20)$ & $0.402(36)$  \\
 0.28 & $0.425(20)$ & $0.403(38)$  \\
 0.41 & $0.374(20)$ & $0.379(41)$  \\
 0.54 & $0.357(28)$ & $0.322(58)$  \\
 0.66 & $0.352(26)$ & $0.427(54)$  \\
\hline\hline
\end{tabular}
\caption{Form factor $G^{8}_A(Q^2)$ extracted from a linear fit of~\eqref{eq:summation} for $t_w^{\mathrm{low}} \leq t_{s,\mathrm{min}} \leq t_w^{\mathrm{up}}$ on J303.}
\label{tab:Q2points_J303}
\end{table}

\begin{table}
\begin{tabular}{l|cc}
\hline\hline
\multicolumn{3}{c}{E300} \\ 
 \hline\hline 
$Q^2$ [GeV$^2$] & \multicolumn{2}{c}{$G^{8}_A(Q^2)$ for $t_{s,\mathrm{min}}$ [fm]} \\ 
  & 0.80 & 0.90 \\
\hline\hline
 0.00 & $0.534(41)$ & $0.483(61)$  \\
 0.07 & $0.507(29)$ & $0.487(48)$  \\
 0.13 & $0.451(28)$ & $0.444(47)$  \\
 0.19 & $0.411(28)$ & $0.431(44)$  \\
 0.25 & $0.402(29)$ & $0.350(46)$  \\
 0.31 & $0.384(27)$ & $0.376(47)$  \\
 0.37 & $0.381(26)$ & $0.395(44)$  \\
 0.48 & $0.340(34)$ & $0.329(63)$  \\
 0.53 & $0.323(28)$ & $0.253(52)$  \\
 0.58 & $0.255(34)$ & $0.202(63)$  \\
 0.63 & $0.293(32)$ & $0.249(60)$  \\
 0.68 & $0.236(44)$ & $0.146(79)$  \\
\hline\hline
\end{tabular}
\caption{Form factor $G^{8}_A(Q^2)$ extracted from a linear fit of~\eqref{eq:summation} for $t_w^{\mathrm{low}} \leq t_{s,\mathrm{min}} \leq t_w^{\mathrm{up}}$ on E300.}
\label{tab:Q2points_E300}
\end{table} 

\end{document}